\documentclass[pra,aps,superscriptaddress,footinbib,twocolumn]{revtex4-1}

\usepackage{mathrsfs}
\usepackage{amsfonts}
\usepackage{amssymb}
\usepackage{amsmath}
\usepackage{graphicx}
\usepackage[usenames,dvipsnames]{color}
\usepackage[colorlinks=true,citecolor=blue,linkcolor=magenta]{hyperref}
\usepackage{ulem}
\usepackage{lmodern}
\usepackage{amsthm}
\usepackage{dsfont, natbib}
\usepackage{dcolumn}

\AtBeginDocument{%
    \newwrite\bibnotes
    \def\bibnotesext{Notes.bib}
    \immediate\openout\bibnotes=\jobname\bibnotesext
    \immediate\write\bibnotes{@CONTROL{REVTEX41Control}}
    \immediate\write\bibnotes{@CONTROL{%
    apsrev41Control,author="08",editor="1",pages="1",title="1",year="1"}}
     \if@filesw
     \immediate\write\@auxout{\string\citation{apsrev41Control}}%
    \fi
}%

\newcommand{\figpath}{.}

\newcommand{\Tr}{\mathrm{Tr}}
\newcommand{\norm}[1]{\Vert #1 \Vert}
\newcommand{\abs}[1]{\vert #1 \vert}

\newcommand{\ket}[1]{\vert{ #1 }\rangle}

\newcommand{\ketbra}[2]{\vert #1 \rangle \langle #2 \vert}

\newcommand{\mean}[1]{\langle #1 \rangle}

\newcommand{\Error}{\mathrm{Error}}
\newcommand{\com}{\mathrm{com}}
\newcommand{\comef}{\mathrm{com}^{\rm ef}}
\newcommand{\Fidelity}{\mathrm{Fidelity}}
\newcommand{\Pro}{\mathrm{Pro}}
\newcommand{\Proef}{\mathrm{Pro}^{\rm ef}}
\newcommand{\Var}[1]{\mathrm{Var}\left(#1\right)}

\newcommand{\bbR}{\mathbb{R}}
\newcommand{\bbU}{\mathbb{U}}
\newcommand{\bbC}{\mathbb{C}}
\newcommand{\bbH}{\mathbb{H}}

\newcommand{\bfF}{\boldsymbol{F}}
\newcommand{\bfR}{\boldsymbol{R}}
\newcommand{\bfmu}{\boldsymbol{\mu}}
\newcommand{\bfa}{\boldsymbol{a}}
\newcommand{\bfb}{\boldsymbol{b}}
\newcommand{\bfc}{\boldsymbol{c}}
\newcommand{\bfd}{\boldsymbol{d}}

\newcommand{\bfP}{\boldsymbol{P}}

\newcommand{\calM}{\mathcal{M}}

\newcommand{\calJ}{\mathcal{J}}
\newcommand{\calK}{\mathcal{K}}
\newcommand{\calMef}{\mathcal{M}^{\rm ef}}
\newcommand{\calGef}{\mathcal{G}^{\rm ef}}
\newcommand{\calRef}{\mathcal{R}^{\rm ef}}
\newcommand{\calI}{\mathcal{I}}
\newcommand{\calB}{\mathcal{B}}
\newcommand{\calN}{\mathcal{N}}

\begin{document}

\title{Scalable evaluation of quantum-circuit error loss using Clifford sampling}

\affiliation{Interdisciplinary Center for Quantum Information, State Key Laboratory  of Modern Optical Instrumentation,
and Zhejiang Province Key Laboratory of Quantum Technology and Device,\\
Department of Physics, Zhejiang University, Hangzhou 310027, China\\
$^2$C. N. Yang Institute for Theoretical Physics, State University of New York at Stony Brook, \\
Stony Brook, NY 11794-3840, USA\\
$^3$Department of Physics and Astronomy, State University of New York at Stony Brook, \\
Stony Brook, NY 11794-3800, USA\\
$^4$Graduate School of China Academy of Engineering Physics, Beijing 100193, China\\}

\author{Zhen Wang$^{1, *}$}
\author{Yanzhu Chen$^{2, 3, *}$}
\author{Zixuan Song$^{1}$}
\author{Dayue Qin$^{4}$}
\author{Hekang Li$^{1}$}
\author{Qiujiang Guo$^{1}$}
\author{H. Wang$^{1}$}
\author{Chao Song$^{1, \dagger}$}
\author{Ying Li$^{4, \ddagger}$}

\begin{abstract}
A major challenge in developing quantum computing technologies is to accomplish high precision tasks by utilizing multiplex optimization approaches, on both the physical system and algorithm levels. Loss functions assessing the overall performance of quantum circuits can provide the foundation for many optimization techniques. In this paper, we use the quadratic error loss and the final-state fidelity loss to characterize quantum circuits. We find that the distribution of computation error is approximately Gaussian, which in turn justifies the quadratic error loss. It is shown that these loss functions can be efficiently evaluated in a scalable way by sampling from Clifford-dominated circuits. We demonstrate the results by numerically simulating ten-qubit noisy quantum circuits with various error models as well as executing four-qubit circuits with up to ten layers of two-qubit gates on a superconducting quantum processor. Our results pave the way towards the optimization-based quantum device and algorithm design in the intermediate-scale quantum regime. 
\end{abstract}

\maketitle

\section{Introduction}

In quantum computation, errors caused by decoherence and imperfect controls form the main obstacle to meaningful applications, such as solving integer factorization and quantum chemistry problems~\cite{Nielsen2010, Shor1994, Peruzzo2014, McArdle2020}. Evaluating the error severity in quantum computation is essential for improving the design of device~\cite{2009_PRL_RBandQPT, 2011_NatPhy_NoiseSpec, 2020_PRAp_OptimizeSQUID}, optimizing control parameters~\cite{2014_PRL_CRB}, and minimizing errors with mitigation protocols~\cite{Li2017, Temme2017, Endo2018}. Various schemes of quantum system characterization have been developed. Randomized benchmarking~\cite{Emerson2005, Knill2008, Dankert2009, Magesan2011, Epstein2014, Kimmel2014, 2015_PRL_AverageRB, Roth2018, 2019_PRL_DRB, 2019_PRL_3QRB} and quantum process tomography (QPT)~\cite{Chuang1997, Poyatos1997, DAriano2001, Altepeter2003, Mohseni2006,  BlumeKohout2017} can measure the average gate fidelity and full information of a noisy quantum channel, respectively. These two methods are efficient in systems with a few qubits. Cross-entropy benchmarking is used to verify a multi-qubit system but cannot be directly applied to quantum-supremacy circuits that are unsimulatable on classical computers~\cite{Boixo2018, Arute2019}. We can infer the performance of a large system by dividing it into tractable subsystems and characterizing each subsystem individually~\cite{Endo2018, Arute2019, Govia2020, Cotler2020, Geller2020, Hamilton2020}. However, this approach only works when the crosstalk is insignificant. The temporal correlation of noise is another factor that usually limits the effectiveness of characterization techniques~\cite{Wallman2014, Fogarty2015, Ball2016, BlumeKohout2017, Mavadia2018, Rudinger2019, Veitia2018, Huo2018}. 

Many quantum algorithms utilize multi-qubit and deep quantum circuits. Even for variational quantum computation, which is promising for near-term applications, we need to implement hundreds of gates on tens of qubits~\cite{Wecker2014, Moll2016, Moll2018, DallaireDemers2019, Gard2020}. In this paper, we propose an intuitive method that can efficiently characterize large quantum circuits, in the presence of both spatial and temporal error correlations. The resource cost of our method scales polynomially with the circuit size. 

\begin{figure}[tbp]
\begin{center}
\includegraphics[width=1\linewidth]{\figpath/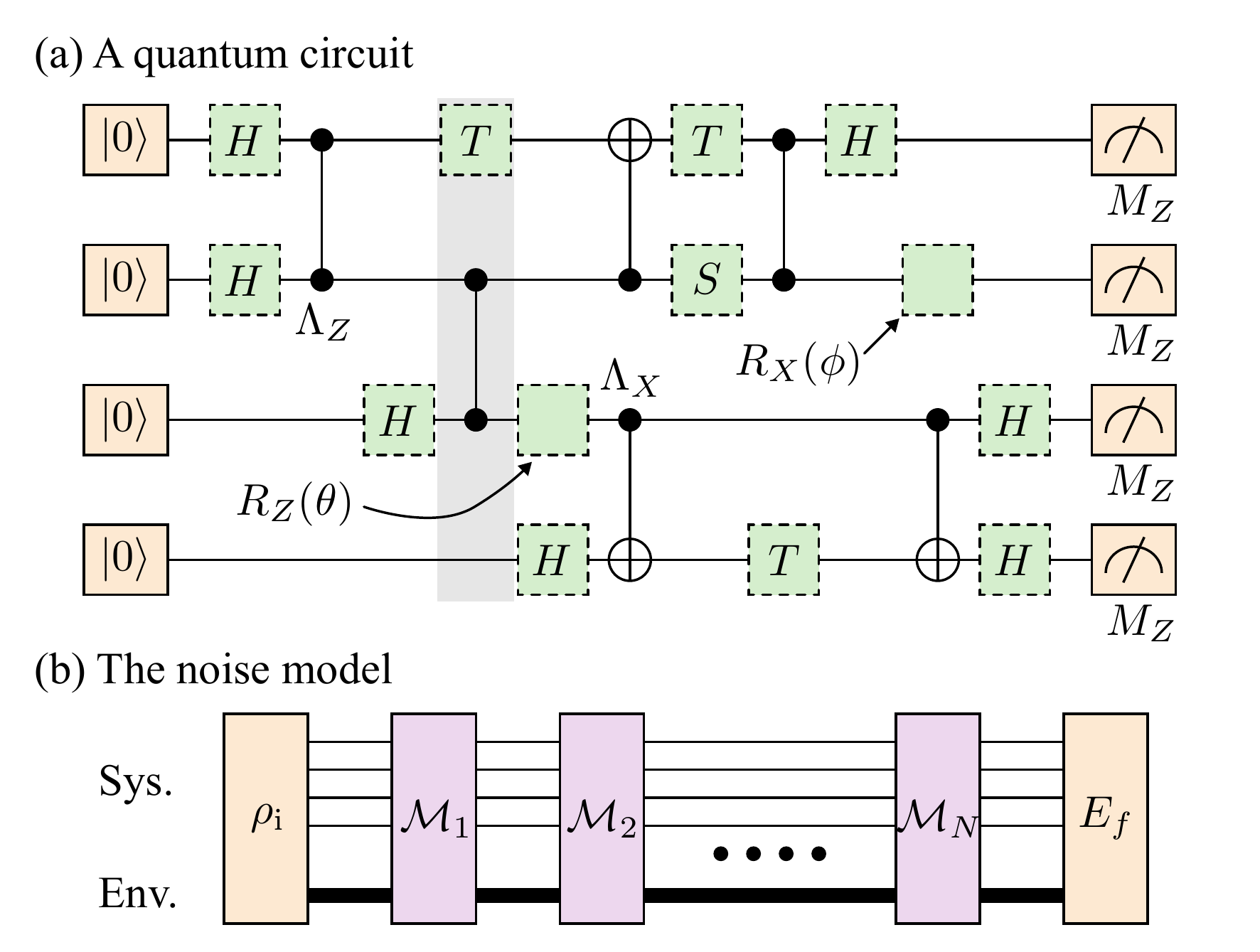}
\caption{
(a) A quantum circuit. The circuit $(\bfF,\bfR)$ consists of frame operations $\bfF$ and single-qubit unitary gates $\bfR$. Single-qubit gates are the green dashed squares, and all other operations are frame operations. In the random circuit sampling for evaluating $L_{\bbR}(\bfF)$, frame operations are always the same, but we alter single-qubit gates to different configurations. 
(b) The noise model. The system (Sys.) formed by qubits and the environment (Env.) are initialized in the state $\rho_{\rm i}$. Following the initialization, there is a sequence of completely-positive maps applied. The map $\mathcal{M}_\tau$ describes the evolution applied at the time $\tau$ to realize a layer (column) of quantum gates, which actually acts on the system and environment because of imperfections. Finally, we implement the measurement on each qubit. The operator of observable is $E_f$, which acts on both the system and environment because of imperfections. 
}
\label{fig:circuit}
\end{center}
\end{figure}

We take the quadratic loss function of computation error~\cite{Strikis2020} as the measure of error severity, which is 
\begin{eqnarray}
L_{\bbR}(\bfF) \equiv \frac{1}{\abs{\bbR}} \sum_{\bfR\in \bbR} \Error(\bfF,\bfR)^2.
\end{eqnarray}
Here $\Error(\bfF,\bfR) \equiv \com(\bfF,\bfR) - \comef(\bfF,\bfR)$ is the computation error, $\com(\bfF,\bfR)$ and $\comef(\bfF,\bfR)$ are respectively results (means of an observable) in the actual noisy computation and error-free computation, and $(\bfF,\bfR)$ specifies a quantum circuit. This loss function characterizes errors in a set of circuits with the same frame operations $\bfF$ as shown in Fig.~\ref{fig:circuit}(a): Frame operations include the qubit initialization, measurement and multi-qubit entangling gates (e.g.~controlled-NOT and controlled-phase gates), which are usually error-prone compared with single-qubit gates. We focus on the case that entangling gates are all Clifford. Single-qubit gates denoted by $\bfR$ are different in the set of circuits. $\bbR$ is the set of single-qubit gate configurations. When $\bbR = \bbU$, single-qubit gates can be any unitary transformations, and the summation should be taken as integration with respect to Haar measure; when $\bbR = \bbC$, single-qubit gates are all Clifford. Taking single-qubit gates as variables is a natural way to construct ansatz circuits in variational quantum algorithms~\cite{Peruzzo2014, Kandala2017, Havlicek2019}. Loss functions in this form can be used to determine parameters in the learning-based quantum error mitigation~\cite{Strikis2020,Czarnik2020,Cincio2020}. 

In this paper, we demonstrate that the quadratic error loss $L_{\bbU}(\bfF)$ (i.e.~$\bbR = \bbU$) is a good objective function and can be efficiently evaluated when the circuit is large. By sampling random circuits, we study the statistics of $\Error(\bfF,\bfR)$ in experiments on a quantum device with four superconducting qubits and numerical simulations with up to ten qubits using various error models. We find that the error distribution is approximately Gaussian with zero mean when general unitary circuits are uniformly sampled from $\bfR\in \bbU$ according to Haar measure, i.e.~$L_{\bbU}(\bfF)$ is the only value that we need for characterizing the error statistics~\cite{footnote}. Computing the error-free result $\comef(\bfF,\bfR)$ is impractical for large general unitary circuits but efficient for Clifford circuits, according to the Gottesman-Knill theorem~\cite{Gottesman1998, Nielsen2010}. We prove $L_{\bbU}(\bfF) = L_{\bbC}(\bfF)$, i.e.~we can obtain $L_{\bbU}(\bfF)$ by only sampling Clifford circuits, under the assumption that errors in single-qubit gates are gate-independent. Note that we do not need any assumptions on frame-operation errors. The equivalence between error losses of unitary sampling and Clifford sampling is verified in both experiments and numerical simulations. 

\section{Fidelity loss and hybrid sampling}

In addition to the quadratic error loss, the final-state fidelity loss $E_{\bbU}(\bfF)$ can also be efficiently evaluated using Clifford sampling. The fidelity loss reads (see Appendix~\ref{app:fidelity}) 
\begin{eqnarray}
E_{\bbR}(\bfF) \equiv \frac{1}{\abs{\bbR}} \sum_{\bfR\in \bbR} [1-\Fidelity(\bfF,\bfR)]. 
\end{eqnarray}
The fidelity loss measures the overall quality of final states, compared with the quadratic error loss defined for specific computation tasks (observables). 

In the fully-Clifford sampling, we assume single-qubit-gate errors are gate-independent. The weak gate dependence can be accounted for by hybridizing Clifford circuits with a few general unitary single-qubit gates. We remark that Clifford-dominated circuits can be efficiently simulated using classical computer~\cite{Bravyi2016}. The fidelity loss and hybrid sampling are studied analytically and numerically in Appendix \ref{app:fidelity} and \ref{app:hybrid}. In the following, we focus on the quadratic error loss and fully-Clifford sampling. 

\section{Formalism and Clifford sampling}

In a quantum circuit, we can draw gates applied in parallel in the same layer (column), see Fig.~\ref{fig:circuit}(a). For example, the gray box is the fourth layer, which contains a $T$ gate and a controlled-phase gate. When gates are error-free, the overall map of the fourth layer is $\calMef_4 = [T \otimes \Lambda_Z \otimes I]$, where $[U](\rho) = U\rho U^\dag$, and $I$ is the identity operator of a qubit. The gates are realized through time evolution. Because of the noise, the actual time evolution leads to a different map $\calM_4$, which acts on not only qubits but also the environment. This is a general formalism of errors in the quantum computation, including both spatial and temporal correlations. The temporal correlation is caused by the environment. According to this formalism, we can express the actual computation result with error as $\com(\bfF,\bfR) = \Tr[E_f \calM_N\cdots\calM_2\calM_1(\rho_{\rm i})]$ for an $N$-layer circuit [see Fig.~\ref{fig:circuit}(b)]. Here, $\rho_{\rm i}$, $E_f$ and $\calM_\tau$ depend on $\bfF$ and $\bfR$ (Temporal correlations can cause the dependence). Qubits are measured in the computational basis, and the outcome is a binary vector $\bfmu$. The corresponding measurement operator is $E_{\bfmu}$. We consider the case that the computation result is the mean of a real function $f(\bfmu)$, then $E_f = \sum_{\bfmu} f(\bfmu) E_{\bfmu}$. 

We can express the error-free map $\calMef_\tau$ as a product of frame gates $\calGef_\tau$ and single-qubit gates $\calRef_\tau$, i.e.~$\calMef_\tau = \calGef_\tau \calRef_\tau$. For example, we have $\calGef_4 = [I \otimes \Lambda_Z \otimes I]$ and $\calRef_4 = [T \otimes I^{\otimes 3}]$. The actual map can always be expressed in the form $\calM_\tau = \mathcal{J}_\tau \left(\calRef_\tau\otimes[\openone_{\rm E}]\right) \mathcal{K}_\tau$. Here, $\openone_{\rm E}$ is the identity operator of the environment, and $\mathcal{J}_\tau$ and $\mathcal{K}_\tau$ are maps on both the system and environment. $\calM_\tau$ in this form is a linear map for matrix entries of $\calRef_\tau$. Therefore, we have the tensor form of the quantum computation $\com(\bfF,\bfR) = \Tr[(\overline{R}\otimes \overline{R}^*)F]$, where $\overline{R}$ is the tensor product of error-free single-qubit gates [e.g.~$\overline{R}=H^{\otimes 3}\otimes T\otimes R_Z(\theta)\otimes\cdots$ in Fig.~\ref{fig:circuit}(a), in which gates are listed from top to bottom then left to right], and $F$ is a tensor describing the effect of frame operations (see Appendix~\ref{app:TensorRep}). Errors are single-qubit-gate-independent (i.e.~$\bfR$-independent) if $\rho_{\rm i}$, $E_f$, $\mathcal{J}_\tau$ and $\mathcal{K}_\tau$ (for all $\tau$) are constants. Then ${\rm Error} (\bfF,\bfR)^2$ is a homogeneous polynomial of degree $2$ in both matrix elements of single-qubit gates and their Hermitian conjugates. The Clifford group is a unitary $2$-design~\cite{Gross2007, Roy2009, Dankert2009}, and therefore $L_{\bbU}(\bfF) = L_{\bbC}(\bfF)$. We remark that, not only the second- but also the first- and third-order moments of the error distribution in unitary sampling can be evaluated using the Clifford sampling, because the Clifford group is also a $1$-design and $3$-design~\cite{Webb2015, Zhu2017}. 

In the Clifford sampling, we uniformly sample each single-qubit gate in the circuit from the Clifford group. We compute the error loss using the Monte Carlo summation method. There are two approaches. In the mean-value approach, we run each random circuit for multiple times on the actual quantum computer and in the simulation on a classical computer to estimate $\com(\bfF,\bfR)$ and $\comef(\bfF,\bfR)$, respectively. Then, we can compute the error loss directly according to its definition. This approach is used in our experiments and numerical simulations. In the single-run approach, each random circuit only runs for once or twice. Then, the variance (due to finite sampling) of the $L_{\bbC}(\bfF)$ estimator is upper bounded by $4\norm{E_f}^4/N_{\rm s}$, and $4N_{\rm s}$ circuit runs are implemented on both quantum and classical computers (see Appendix~\ref{app:MC}). Our method is scalable since the variance is independent of the circuit depth and the number of qubits. 

\begin{figure}[tbp]
\centering
\includegraphics[width=1\linewidth]{\figpath/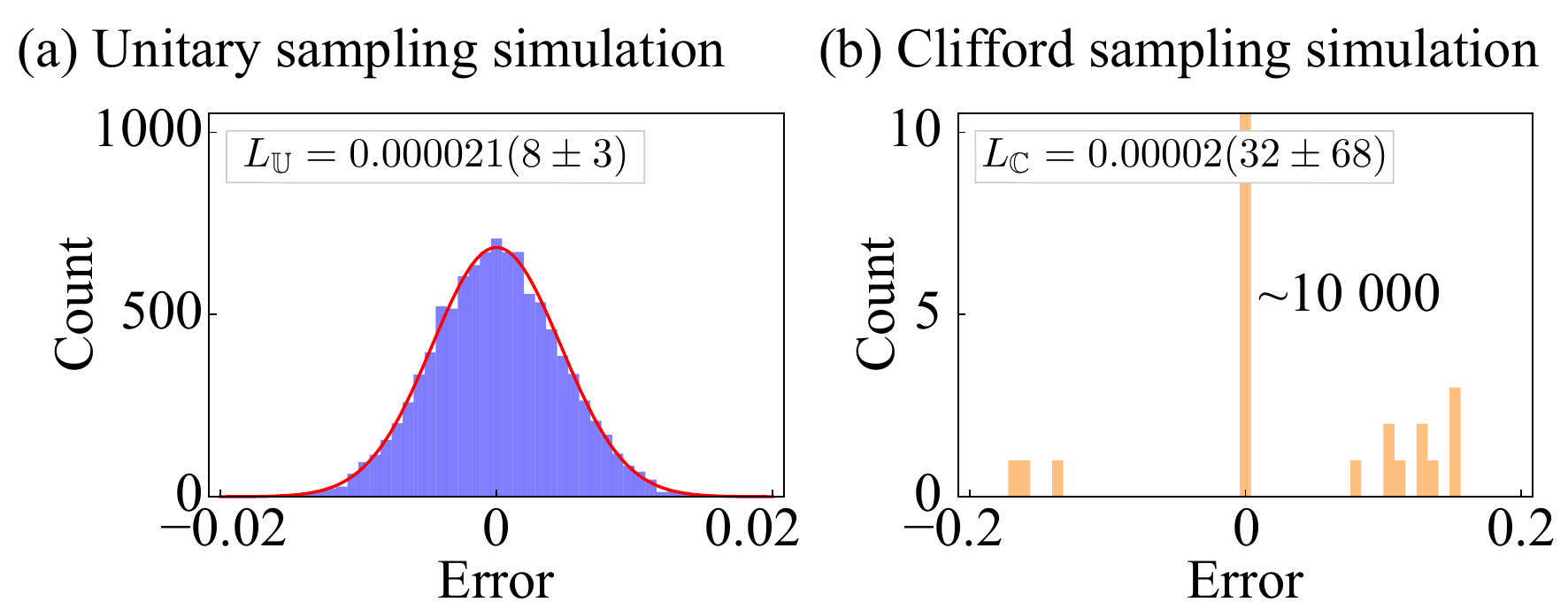}
\caption{
Numerical results. (a) The unitary sampling of $\Error(\bfF,\bfR)$. The red curve denotes the Gaussian distribution $\mathcal{N}(0,L_{\bbU})$. (b) The Clifford sampling of $\Error(\bfF,\bfR)$. A ten-qubit circuit with a hundred two-qubit gates is used to compute the mean of a Pauli operator. 
}
\label{fig:DepolMain}
\end{figure}

\section{Numerical results}

We implement numerical simulations of quantum circuits using the library QuESTlink~\cite{Jones2020, Jones2019} for various error models, frame operation configurations and up to ten qubits. Error models include the depolarizing, dephasing, amplitude damping, correlated coherent, gate-dependent depolarizing, composite and experimentally-measured models. Here, we only show the results of ten qubits with the depolarizing model and a specific frame operation configuration. See Appendix~\ref{app:numerics} for details and results of other error models. 

In Fig.~\ref{fig:DepolMain}(a), we plot the distribution of $\Error(\bfF,\bfR)$ in the unitary sampling, and we can find that the distribution is approximately Gaussian. This conclusion holds in all our numerical simulations and experiments, for various circuit sizes, error models and frame operation configurations. See Appendix~\ref{app:numerics} for a comparison between moments of $\Error(\bfF,\bfR)$ and the Gaussian distribution. 

The error distribution is non-Gaussian in the Clifford sampling, as shown in Fig.~\ref{fig:DepolMain}(b). In all simple error models (i.e.~depolarizing, dephasing, amplitude damping, and coherent models), the distribution is discretized and concentrated at several values of the error, and most of the probability is concentrated at zero. We can understand this result as follows~\cite{Strikis2020}. For Clifford circuits, if the observable to be measured $E_f$ is a Pauli operator as in our case, $\comef(\bfF,\bfR)$ takes three values $0$ or $\pm 1$. For most of the cases, $\comef(\bfF,\bfR) = 0$, and we always have $\com(\bfF,\bfR) = 0$ if errors are Pauli, i.e.~$\Error(\bfF,\bfR) = 0$. Therefore, for Pauli and Pauli-like errors, many Clifford circuits are error-insensitive. We can improve the efficiency of evaluating $L_{\bbC}(\bfF)$ using the importance sampling by selecting error-sensitive Clifford circuits. 

\begin{figure}[tbp]
\centering
\includegraphics[width=1\linewidth]{\figpath/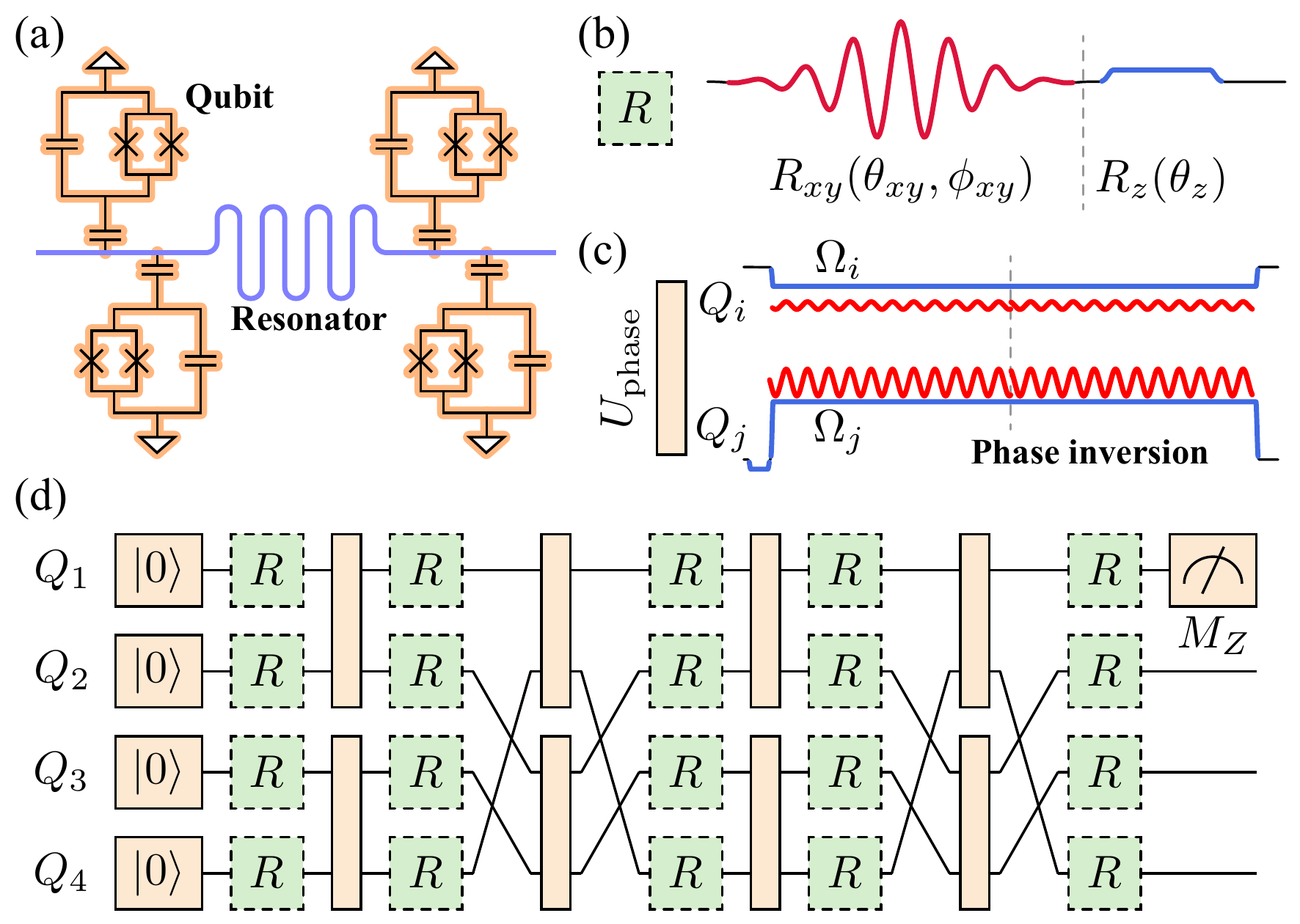}
\caption{
Experiment setup. (a) Diagram of the superconducting qubit device. We use four Xmon qubits in the experiment. All  of them are coupled to a central bus resonator. (b) A single-qubit gate is realized using a $R_{xy}$ gate followed by a $R_{z}$ gate. The $R_{xy}$ gate has two parameters $\theta_{xy}$ and $\phi_{xy}$, which are controlled by the amplitude and phase of XY pulse (red). The pulse length is 40 ns. The $R_{z}$ gate has only one parameter $\theta_{z}$, which is controlled by the amplitude of the Z pulse (blue), whose length is 10 ns. (c) A two-qubit gate $U_{\text{phase}}$ is generated by tuning two qubits into near resonance (long square Z pulse in blue) with sinusoidal microwave driving field (red) being applied on each qubit dynamically. A short Z pulse before the long Z pulse is applied on one of the qubits to align the $x$-axes of their Bloch spheres. We note that the driving field ($\Omega_{i}$ on the qubit $Q_{i}$) is different in $U_{\text{phase}}$ gates on different qubits. The length of $U_{\text{phase}}$ is around 300 ns. Phases of driving fields are inverted at the middle of the gate. (d) The quantum circuit used to demonstrate the Clifford sampling, in which we fix two-qubit gates and left single-qubit gates as variables. 
}
\label{fig:Sequence}
\end{figure}

\section{Experimental results}

To demonstrate  the feasibility and usefulness of Clifford sampling in an actual quantum computer, we implement it on a superconducting quantum device, which is illustrated in Fig.~\ref{fig:Sequence}(a). Four frequency-tunable Xmon qubits ($Q_1 \sim Q_4$) are coupled to a central bus resonator, which mediates the effective interaction between qubits for implementing two-qubit gates. For every single-qubit gate $R\in{\rm U}(2)$, we can  decompose it into two experimentally feasible gates $R = e^{i\alpha}R_{z}(\theta_z)R_{xy}(\theta_{xy}, \phi_{xy})$, as shown in Fig.~\ref{fig:Sequence}(b), where $\alpha$, $\theta_{z}$, $\theta_{xy}$ and $\phi_{xy}$ are real numbers. For two-qubit gates, we use the Clifford dressed-state gate $U_{\text{phase}}$, which is essentially the controlled-phase gate but in the $X$ basis [see Fig.~\ref{fig:Sequence}(c)]~\cite{Guo2018}. We can implement the gate $U_{\text{phase}}$ between any pair of qubits, therefore we have six gate setups for four qubits. See Appendix~\ref{app:exp} for device parameters and detailed implementation of gates.

Before applying Clifford sampling, we benchmarked fidelities of single-qubit gates and two-qubit gates. Fidelities of six $U_{\text{phase}}$ setups measured using QPT are $95\%\sim 97\%$. In some circuits, two $U_{\text{phase}}$ are applied in parallel. Because of crosstalk, gate fidelities are changed slightly in parallel operations. We use randomized benchmarking to measure gate fidelities of $R_{xy}$ and $R_{z}$, which is implemented on each qubit individually as well as simultaneously on all qubits. Both approaches yield no less than $99.3\%$ fidelities for $X$, $Y$, $X/2$ and $Z$ gates. The average error rate of single-qubit gates is at least an order of magnitude lower than two-qubit gates, thus we can safely infer that most of the noise is introduced by $U_{\text{phase}}$. The gate performance can be improved by optimization based on Clifford sampling. See Appendix~\ref{app:exp} for benchmarking and optimization data. 

We use the circuit in Fig.~\ref{fig:Sequence}(d) as an example to implement the Clifford sampling. The observable to be measured is the probability of $Q_1$ being in $\ket{0}$, i.e.~$E_f = \ketbra{0}{0}_1 = (I_1+Z_1)/2$. Given a specific circuit $(\bfF,\bfR)$, we run the circuit for 1000 times in order to estimate the probability in $\ket{0}$. The probability obtained in the experiment is $P_0^{\rm exp}$, and its error-free value computed using the classical computer is $P_0^{\rm ef}$. We note that $P_0^{\rm exp}$ has been corrected for readout errors (see Appendix~\ref{app:exp}). Then, the computation error is $\Error = P_0^{\rm exp} - P_0^{\rm ef}$. 

\begin{figure}[tbp]
\centering
\includegraphics[width=1\linewidth]{\figpath/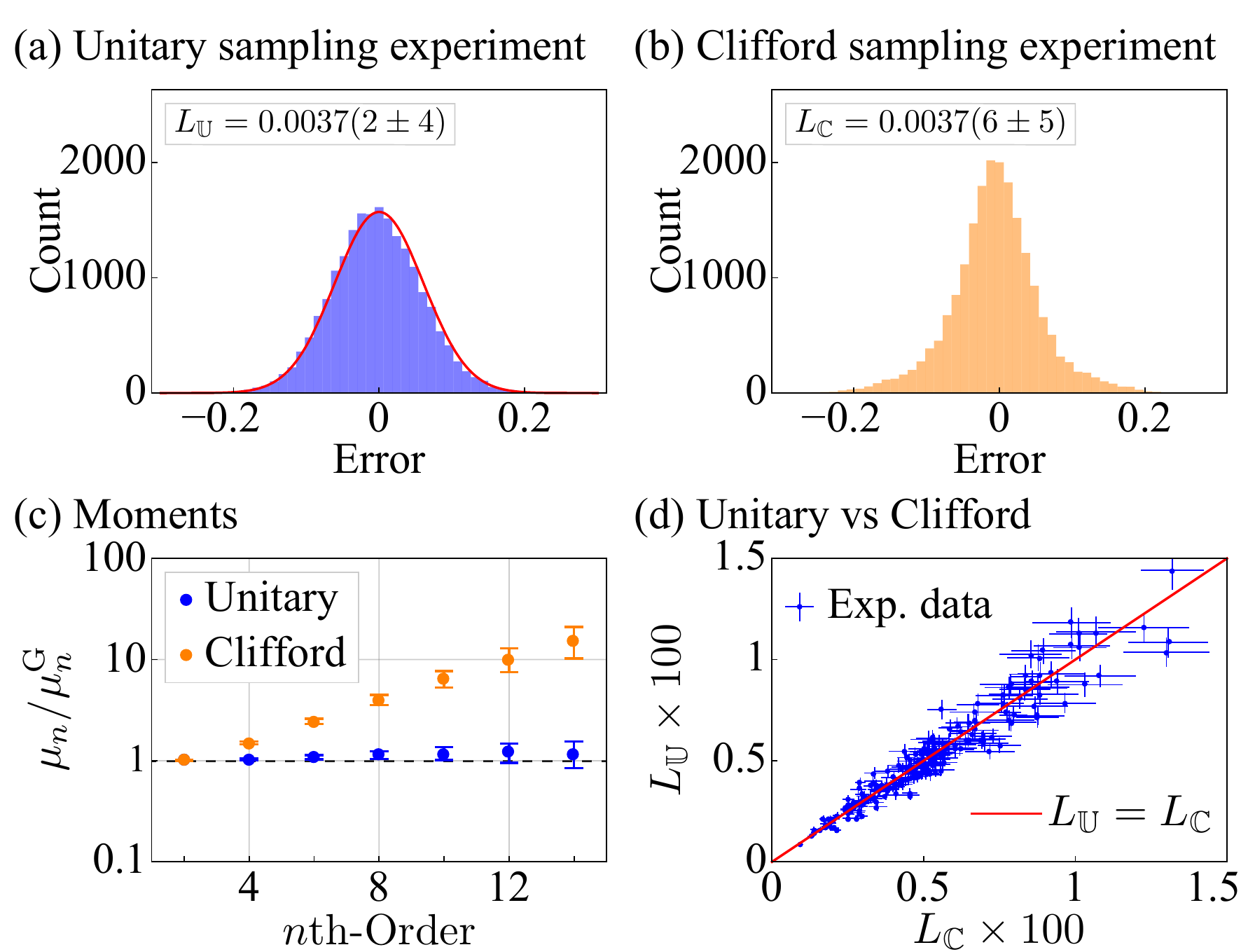}
\caption{
Experimental results. (a) The unitary sampling of $\Error(\bfF,\bfR)$. The red curve denotes the Gaussian distribution $\mathcal{N}(0,L_{\bbU})$. (b) The Clifford sampling of $\Error(\bfF,\bfR)$. (c) Moments of two sampling approaches. $\mu_n = {\rm E}[\Error^n]$ is the $n$th-order moment, and $\mu_n^{\rm G}$ is the moment of $\mathcal{N}(0,L_{\bbU})$. The quantum circuit used to generate data in (a), (b) and (c) is in Fig.~\ref{fig:Sequence}(d). (d) Error losses of unitary sampling versus Clifford sampling for 50 randomly generated frame operation configurations with one to ten layers of two-qubit gates. Each layer has one or two $U_{\text{phase}}$ gates. When only one gate is applied, we protect the other two qubits from dephasing by applying dynamically drive fields~\cite{Guo2018}. The observable is $\ketbra{0}{0}$ for one of four qubits. See Appendix~\ref{app:exp} for details. The error bar denotes the standard deviation due to finite sampling. 
}
\label{fig:EXP}
\end{figure}

Both unitary sampling and Clifford sampling are implemented in the experiment. For each case, $20 000$ random configurations of single-qubit gates $\bfR$ are generated. In the unitary sampling, the error distribution is Gaussian as shown in Fig.~\ref{fig:EXP}(a), the same as in numerical simulations. However, in the Clifford sampling, the error distribution is continuous as shown in Fig.~\ref{fig:EXP}(b), which is obviously different from the numerical results of simple error models. In Appendix~\ref{app:numerics}, we give numerical results of a composite error model (a combination of coherent and amplitude damping errors) and the experimentally-measured model (from QPT). The error distribution in Clifford sampling for these two models are continuous and in qualitative agreement with the experimental result. We plot moments up to the $14$th-order in Fig.~\ref{fig:EXP}(c): Moments of the unitary sampling are consistent with the Gaussian distribution, and moments of the Clifford sampling converge more slowly than Gaussian. Although two distributions are different, their 2nd-order moments, i.e.~the loss function values $L_{\bbU}=0.0037(2\pm 4)$ and $L_{\bbC}=0.0037(6\pm 5)$, are the same up to the sampling noise. 

In addition to the circuit in Fig.~\ref{fig:Sequence}(d), we implemented the experiment for 50 randomly generated frame operation configurations $\bfF$. The error loss ($L_{\bbU}$ or $L_{\bbC}$) is estimated by sampling 500 single-qubit gate configurations $\bfR$ for each frame operation configuration. The result is plotted in Fig.~\ref{fig:EXP}(d). Almost all data points are within $2\sigma$ from the diagonal line, which represents $L_{\bbU}=L_{\bbC}$. $\bfR$-dependent errors can cause the difference between two error losses (which is observed in numerical simulations of the gate-dependent error model), however, this effect is not significant in the experimental result. 

\section{Discussion}

We propose to characterize quantum circuits executed on a noisy device by evaluating the quadratic error loss and fidelity loss using the Clifford sampling method. In these two loss functions, all the temporal and spatial correlations are automatically taken into account by treating the entire circuit as a whole. We demonstrate the Clifford sampling method with both numerical simulations and experiments on a superconducting device. We prove that fully-Clifford sampling is sufficient as long as the noise is independent of single-qubit gates. Weak gate dependence can be tackled using hybrid sampling. Experimental results do not show the significant effect of gate dependence. We observe a continuous distribution of the computation error in Clifford sampling in the experiment, whereas some simple error models such as Pauli error models predict a discretized distribution. This result suggests that these models cannot correctly describe the noise in our experiment. One can verify an error model and determine its parameters by constructing loss functions to compare the actual device with the error model. 

In addition to characterizing quantum circuits, our method can find application in optimizing their performance. We experimentally implemented the optimization of a Rabi frequency driving two-qubit gates in a set of four-qubit four-depth circuits. The error losses decrease by more than $10\%$ by using Clifford sampling (see Appendix~\ref{app:exp}). In the single-run approach, the sampling cost does not scale with the circuit size. Therefore, our method is promising in the multi-parameter optimization for large-scale quantum circuits. Other than optimizing parameters, our method can provide ground for choosing circuits. The circuit for a computation task may not be unique. Given a noisy quantum device, one can select a working circuit among theoretically equivalent circuits based on our loss functions. A similar idea was proposed in Ref.~\cite{Cincio2020}. Compared to assessing the general performance of the device, our scheme is more application-oriented, i.e.~each characterization experiment reflects the likelihood that the device performs well in solving a particular problem. Consequently, the optimization based on our characterization is tailored for specific problems and corresponding circuits. 

\begin{acknowledgments}
We thank Wuxing Liu and Xu Zhang for discussions and technical support. YC thanks Tzu-Chieh Wei and Jason P. Kestner for insightful discussions. We are grateful to the authors of QuESTlink for making their package public. We acknowledge the support of the National Natural Science Foundation of China (No. 11725419 and No. 11875050), the National Key Research and Development Program of China (Grants No. 2017YFA0304300, No. 2019YFA0308100), the Zhejiang Province Key Research and Development Program (Grant No. 2020C01019) and the Basic Research Funding of Zhejiang University. YC is supported by the National Science Foundation (Grant No. PHY 1915165) and BNL LDRD \#19-002. DYQ and YL are also supported by NSAF (Grant No. U1930403). 
\end{acknowledgments}

\appendix

\setcounter{figure}{0}
\setcounter{table}{0}
\renewcommand\thefigure{S\arabic{figure}}
\renewcommand\thetable{S\arabic{table}}

\section{Tensor representation of quantum circuits}
\label{app:TensorRep}

Let $\bfR = (R_1,R_2,\ldots,R_{N_R})$ be the list of single-qubit gates, where $R_i\in {\rm U}(2)$ is a single-qubit unitary gate. The $i$th single-qubit gate is applied on the $l_i$th qubit in the $t_i$th layer. The overall map of error-free single-qubit gates in the $\tau$th layer is $\calRef_\tau = [R_{\tau,1}]\otimes[R_{\tau,2}]\otimes\cdots\otimes[R_{\tau,n}]$, where $n$ is the number of qubits, and $R_{\tau,m}$ is the single-qubit gate on the $m$th qubit in the $\tau$th column, which is either one of $R_i$'s in $\bfR$ or identity. We have $R_{\tau,m} = \prod_{i=1}^{N_R} R_i^{\delta_{\tau,t_i}\delta_{m,l_i}}$. The tensor product of all single-qubit gates is $\overline{R} = R_1\otimes R_2\otimes\cdots\otimes R_{N_R}$. 

We express the quantum computation result as 
\begin{eqnarray}
\com(\bfF,\bfR) &=& \Tr[E_f \calM_N\cdots\calM_2\calM_1(\rho_{\rm i})]. 
\end{eqnarray}
We can always write the $\tau$th-column map as $\calM_\tau = \calJ_\tau \left(\calRef_\tau\otimes [\openone_{\rm E}]\right) \calK_\tau$. Because $\calRef_\tau$ is invertible, we can take $\calK_\tau = \calI$, which is the identity map, and $\calJ_\tau=\calM_\tau \left( (\calRef_\tau)^{-1} \otimes [\openone_{\rm E}]\right)$. Then, we have
\begin{eqnarray}
&&\com(\bfF;[R_1],[R_2],\ldots,[R_{N_R}]) \notag \\
&=& \Tr[E_f \calJ_N \calRef_N \calK_N \cdots\calJ_2 \calRef_2 \calK_2 \calJ_1 \calRef_1 \calK_1 (\rho_{\rm i})],~~~~~~
\end{eqnarray}
in which we have replaced $\bfR$ with $([R_1],[R_2],\ldots,[R_{N_R}])$, and they are equivalent. 

A single-qubit map can be decomposed as 
\begin{eqnarray}
[R_i] = \sum_{a,b,c,d=0,1} R_{i;a,b} R^*_{i;c,d} \calB_{a,b}^{c,d},
\end{eqnarray}
where $R_i = \sum_{a,b=0,1} R_{i;a,b}\ketbra{a}{b}_i$ is a single-qubit unitary operator, and $\mathcal{B}_{a,b}^{c,d}(\bullet) = \ketbra{a}{b} \bullet \ketbra{d}{c}$ is the natural basis of single-qubit maps. Because of the linearity of $\com(\bfF;[R_1],[R_2],\ldots,[R_{N_R}])$, we have 
\begin{eqnarray}
&&\com(\bfF;[R_1],[R_2],\ldots,[R_{N_R}]) \notag \\
&=& \sum_{\bfa,\bfb,\bfc,\bfd} \left(\prod_{i=1}^{N_R} R_{i;a_i,b_i} R^*_{i;c_i,d_i}\right) F_{\bfa,\bfb}^{\bfc,\bfd},
\end{eqnarray}
where 
\begin{eqnarray}
F_{\bfa,\bfb}^{\bfc,\bfd} = \com(\bfF;\mathcal{B}_{a_1,b_1}^{c_1,d_1},\mathcal{B}_{a_2,b_2}^{c_2,d_2},\ldots,\mathcal{B}_{a_{N_R},b_{N_R}}^{c_{N_R},d_{N_R}}).
\end{eqnarray}
Here, $\bfa = (a_1,a_2,\ldots,a_{N_R})$, $\bfb = (b_1,b_2,\ldots,b_{N_R})$, $\bfc = (c_1,c_2,\ldots,c_{N_R})$ and $\bfd = (d_1,d_2,\ldots,d_{N_R})$ are binary vectors. Taking 
\begin{eqnarray}
F = \sum_{\bfa,\bfb,\bfc,\bfd} F_{\bfa,\bfb}^{\bfc,\bfd} \ketbra{\bfb,\bfd}{\bfa,\bfc},
\end{eqnarray}
we have 
\begin{eqnarray}
\com(\bfF;[R_1],[R_2],\ldots,[R_{N_R}]) = \Tr[(\overline{R}\otimes \overline{R}^*)F],
\end{eqnarray}
 where 
\begin{eqnarray}
\overline{R}\otimes \overline{R}^* = \sum_{\bfa,\bfb,\bfc,\bfd} \left(\prod_{i=1}^{N_R} R_{i;a_i,b_i} R^*_{i;c_i,d_i}\right) \ketbra{\bfa,\bfc}{\bfb,\bfd}.~~
\end{eqnarray}

\section{Monte Carlo method}
\label{app:MC}

Let $f$ be the measurement outcome of the quantum circuit specified by $(\bfF,\bfR)$, and its distribution is $\Pro(f\vert\bfF,\bfR)$. Then, the computing result, i.e.~the mean value of $f$, reads 
\begin{eqnarray}
\com(\bfF,\bfR) = \sum_f \Pro(f\vert\bfF,\bfR) f.
\end{eqnarray}
Similarly, the error-free computing result can be expressed as 
\begin{eqnarray}
\com^{\rm ef}(\bfF,\bfR) = \sum_f \Proef(f\vert\bfF,\bfR) f.
\end{eqnarray}
In the Monte Carlo summation, the distribution $\Pro(f\vert\bfF,\bfR)$ is realized using the actual quantum computer, and all other distributions, including $\Proef(f\vert\bfF,\bfR)$, are realized on the classical computer. 

We express the loss function as 
\begin{eqnarray}
L_{\bbC}(\bfF) &=& \frac{1}{\abs{\bbC}} \sum_{\bfR,f,f'} [\Pro(f\vert\bfF,\bfR) \Pro(f'\vert\bfF,\bfR) ff' \notag \\
&&- 2 \Pro(f\vert\bfF,\bfR) \Proef(f'\vert\bfF,\bfR) ff' \notag \\
&&+ \Proef(f\vert\bfF,\bfR) \Proef(f'\vert\bfF,\bfR) ff'].
\end{eqnarray}

To compute the first term, we generate $N_{\rm s}$ independent and identically distributed samples $\{(\bfR_i,f_i,f_i')\vert i = 1,2,\ldots,N_{\rm s}\}$ according to the distribution 
\begin{eqnarray}
\Pro(\bfR) \Pro(f\vert\bfF,\bfR) \Pro(f'\vert\bfF,\bfR), \notag
\end{eqnarray}
where 
\begin{eqnarray}
\Pro(\bfR) &=& \frac{1}{\abs{\mathbb{C}}}.
\end{eqnarray}
The estimator of the first term is 
\begin{eqnarray}
\hat{L}_1 = \frac{1}{N_{\rm s}} \sum_{i=1}^{N_{\rm s}} f_i f'_i,
\end{eqnarray}
where $f_i$ and $f'_i$ are two independent experimental outcomes obtained for the $i$-th sampled circuit.
The variance of the estimator is 
\begin{eqnarray}
\Var{\hat{L}_1} = \frac{1}{N_{\rm s}} \Var{f f'}.
\end{eqnarray}
Let $\abs{f}_{\rm max}$ be the maximum value of $\abs{f(\bfmu)}$, we have $f f' \leq \abs{f}_{\rm max}^2$. Therefore, 
\begin{eqnarray}
\Var{\hat{L}_1} \leq \frac{1}{N_{\rm s}} \abs{f}_{\rm max}^4.
\end{eqnarray}

To compute the second term, we generate $2N_{\rm s}$ independent and identically distributed samples $\{(\bfR_i,f_i,f_i')\vert i = 1,2,\ldots,2N_{\rm s}\}$ according to the distribution 
\begin{eqnarray}
\Pro(\bfR) \Pro(f\vert\bfF,\bfR) \Proef(f'\vert\bfF,\bfR). \notag
\end{eqnarray}
The estimator of the second term is 
\begin{eqnarray}
\hat{L}_2 = \frac{1}{2N_{\rm s}} \sum_{i=1}^{2N_{\rm s}} f_i f'_i,
\end{eqnarray}
where $f_i$ and $f'_i$ are one experimental outcome and one simulated outcome for the $i$-th sampled circuit respectively.
The variance of the estimator is 
\begin{eqnarray}
\Var{\hat{L}_2} = \frac{1}{2N_{\rm s}} \Var{f f'} \leq \frac{1}{2N_{\rm s}} \abs{f}_{\rm max}^4.
\end{eqnarray}

To compute the third term, we generate $N_{\rm s}$ independent and identically distributed samples $\{(\bfR_i,f_i,f_i')\vert i = 1,2,\ldots,N_{\rm s}\}$ according to the distribution 
\begin{eqnarray}
\Pro(\bfR) \Proef(f\vert\bfF,\bfR) \Proef(f'\vert\bfF,\bfR). \notag
\end{eqnarray}
The estimator of the third term is 
\begin{eqnarray}
\hat{L}_3 = \frac{1}{N_{\rm s}} \sum_{i=1}^{N_{\rm s}} f_i f'_i,
\end{eqnarray}
where $f_i$ and $f'_i$ are two independent simulated outcomes obtained for the $i$-th sampled circuit.
The variance of the estimator is 
\begin{eqnarray}
\Var{\hat{L}_3} = \frac{1}{N_{\rm s}} \Var{f f'} \leq \frac{1}{N_{\rm s}} \abs{f}_{\rm max}^4.
\end{eqnarray}

The estimator of the error loss is 
\begin{eqnarray}
\hat{L}_{\bbC} = \hat{L}_1 -2\hat{L}_2 + \hat{L}_3.
\end{eqnarray}
The variance of the estimator is 
\begin{eqnarray}
\Var{\hat{L}_{\bbC}} &=& \Var{\hat{L}_1} + 4\Var{\hat{L}_2} + \Var{\hat{L}_3} \notag \\
&&\leq \frac{4}{N_{\rm s}} \abs{f}_{\rm max}^4.
\end{eqnarray}

\section{Fidelity loss}
\label{app:fidelity}

The fidelity loss function is
\begin{eqnarray}
E_{\bbR}(\bfF) \equiv \frac{1}{\abs{\bbR}} \sum_{\bfR\in \bbR} [1-\Fidelity(\bfF,\bfR)],
\end{eqnarray}
where for $\bbR=\bbU$ the summation is understood as integration with respect to Haar measure. Given the final state of the quantum circuit $\rho(\bfF,\bfR)$ and the error-free final state $\rho^{\rm ef}(\bfF,\bfR) = \ketbra{\Psi(\bfF,\bfR)}{\Psi(\bfF,\bfR)}$, the fidelity is
\begin{eqnarray}
\Fidelity(\bfF,\bfR) = \Tr[\rho^{\rm ef}(\bfF,\bfR) \rho(\bfF,\bfR)].
\end{eqnarray}
Since we have 
\begin{eqnarray}
&&\rho(\bfF,\bfR) = \Tr_{\rm E}[\calM_N\cdots\calM_2\calM_1(\rho_{\rm i})] \notag \\
&=& \Tr_{\rm E}[\calJ_N \calRef_N \calK_N \cdots\calJ_2 \calRef_2 \calK_2 \calJ_1 \calRef_1 \calK_1(\rho_{\rm i})],~~~~~
\end{eqnarray}
$\rho(\bfF,\bfR)$ is a linear map for each $R_{i;a_i,b_i} R^*_{i;c_i,d_i}$. It is the same for $\rho^{\rm ef}(\bfF,\bfR)$. When the noise is independent of $\bfR$, $\Fidelity(\bfF,\bfR)$ is ${\rm Hom}(2,2)$, where we adopt the notation of homogeneous polynomials from Ref.~\cite{Roy2009}. Therefore, $E_{\bbU}(\bfF)=E_{\bbC}(\bfF)$ and Clifford sampling is sufficient to produce the result for unitary sampling. Compared with the fidelity loss proposed in Ref.~\cite{Strikis2020}, which has a ${\rm Hom}(4,4)$ term, the application of $E_{\bbR}(\bfF)$ in the learning-based quantum error mitigation may have problem, because the error-mitigated state $\rho$ may not be positive semi-definite.

When the circuit is Clifford, i.e.~$\bfR\in \bbC$, the final state $\ket{\Psi(\bfF,\bfR)}$ is a stabiliser state~\cite{Gottesman1998}. Suppose $S_{\bfF,\bfR}$ is the stabiliser group of the state $\ket{\Psi(\bfF,\bfR)}$, we have (see Ref.~\cite{Strikis2020})
\begin{eqnarray}
\Fidelity(\bfF,\bfR) = \frac{1}{2^n} \sum_{g\in S_{\bfF,\bfR}} \Tr\left[g\rho(\bfF,\bfR)\right].
\end{eqnarray}
By measuring the group elements $g$, which are Pauli operators with $\pm$ signs, we can evaluate the fidelity and then the loss function using the Monte Carlo method. In the practical implementation the measurement error may contribute to the result, which is discussed below.

\subsection*{Measurement error in fidelity loss}

To evaluate the fidelity, we need to measure the group elements $g$, which is realized by using an additional layer of single-qubit gates $\calM_{N+1} = \calJ_{N+1} (\calRef_{N+1;g}\otimes[\openone_{\rm E}]) \calK_{N+1}$ to change the effective measurement basis and then measuring all qubits in the $Z$ basis. Here, $\calRef_{N+1;g}$ depends on the element $g$ to be measured, and $\calJ_{N+1}$ and $\calK_{N+1}$ are constants when errors are $\bfR$-independent. We consider the case that single-qubit gates are error-free, i.e.~$\calM_{N+1} = \calRef_{N+1;g}\otimes[\openone_{\rm E}]$. 

For uncorrelated and balanced measurement errors, the measurement outcome is incorrect with a probability $p$, which is the same for both measurement outcome $0$ and $1$, and the event of measurement error is uncorrelated with other operations. Such measurement errors can be expressed as bit-flip errors occurring before the measurement with the probability $p$. Now, we introduce the single-qubit depolarising map 
\begin{eqnarray}
\calN_1(\epsilon) = (1-\epsilon)[I] + \frac{\epsilon}{3}([X]+[Y]+[Z]).
\end{eqnarray}
Because phase-flip errors do not change measurement outcomes in the $Z$ basis, the uncorrelated and balanced measurement errors with the probability $p$ is equivalent to applying $\calN_1(3p/2)$ before the measurement. Let $p_1,p_2,\ldots,p_N$ be measurement error rates of $n$ qubits, the overall measurement-error map is $\calN_{\rm M} = \calN_1(3p_1/2)\otimes\calN_1(3p_2/2)\otimes\cdots\otimes\calN_1(3p_n/2)\otimes[\openone_{\rm E}]$. Then, the mean of $g$ measured in the experiment is actually 
\begin{eqnarray}
\mean{g}_{\rm actual} = \Tr\left[g_Z \calN_{\rm M} \calRef_{N+1;g} \left(\rho(\bfF,\bfR)\right)\right].
\end{eqnarray}
Here, $g_Z = \calRef_{N+1;g}(g)$ is a tensor product of $Z$ operators, which is directly measured at the end of the circuit. The single-qubit depolarising map commutes with single-qubit unitary maps, then 
\begin{eqnarray}
\mean{g}_{\rm actual} &=& \Tr\left[g_Z \calRef_{N+1;g} \calN_{\rm M} \left(\rho(\bfF,\bfR)\right)\right] \notag \\
&=& \Tr\left[g \calN_{\rm M} \left(\rho(\bfF,\bfR)\right)\right]. 
\end{eqnarray}
Therefore, the Clifford sampling measures the fidelity in the state $\calN_{\rm M} \left(\rho(\bfF,\bfR)\right)$, which includes the effect of measurement errors. We remark that the conditions are i) measurement errors are uncorrelated and balanced, and ii) single-qubit-gate errors are negligible. 

\section{Hybrid sampling}
\label{app:hybrid}

When the noise in the circuit depends on single-qubit gates $\bfR$, Clifford sampling may become insufficient for evaluating the loss functions $L_\bbU$ and $E_\bbU$. We show that the hybrid sampling method provides an estimator that can tolerate weak $\bfR$-dependence. Let $F^{\rm ef}$ be the error-free frame-operation tensor, then 
\begin{eqnarray}
L_{\bbR} = \frac{1}{\abs{\bbR}} \sum_{\bfR\in \bbR} \Tr[(\overline{R}\otimes \overline{R}^*)(F-F^{\rm ef})]^2.
\end{eqnarray}
For circuits with $N_R$ single-qubit gates, we can expand $F$ as
\begin{eqnarray}
F = F^{(0)} + \sum_{i=1}^{N_R} F^{(1)}_i(R_i) + \Delta F,
\end{eqnarray}
where $F^{(0)}$ is a constant, $F^{(1)}_i(R_i)$ only depends on the $i$th single-qubit gate $R_i$, and $\Delta F$ is small when the gate-dependence is weak.The expansion of error loss is 
\begin{eqnarray}
L_{\bbR} = L_{\bbR}^{(0)} + 2\sum_{i=1}^{N_R} L_{\bbR;i}^{(1)} + O(\norm{F^{(1)}_i}^2,\norm{\Delta F}),
\end{eqnarray}
where 
\begin{eqnarray}
L_{\bbR}^{(0)} &=& \frac{1}{\abs{\bbR}} \sum_{\bfR\in \bbR} W^{(0)},
\end{eqnarray}
and
\begin{eqnarray}
L_{\bbR;i}^{(1)} &=& \frac{1}{\abs{\bbR}} \sum_{\bfR\in \bbR} W^{(1)}_i.
\end{eqnarray}
Here, 
\begin{eqnarray}
W^{(0)} = \Tr[(\overline{R}\otimes \overline{R}^*)(F^{(0)}-F^{\rm ef})]^2
\end{eqnarray}
is ${\rm Hom}(2,2)$ for all $R_j$, and 
\begin{eqnarray}
W^{(1)}_i &=& \Tr[(\overline{R}\otimes \overline{R}^*)(F^{(0)}-F^{\rm ef})] \notag \\
&&\times \Tr[(\overline{R}\otimes \overline{R}^*)F^{(1)}_i(R_i)]
\end{eqnarray}
is ${\rm Hom}(2,2)$ for all $R_{j\neq i}$ but not for $R_i$. We denote the hybrid sampling set by $\bbH_i\equiv\{R_{j\neq i}\in{\rm C}(2),\,R_i\in{\rm U}(2)\}$. Therefore, we have $L_{\bbU}^{(0)} = L_{\bbC}^{(0)} = L_{\bbH_i}^{(0)}$, $L_{\bbU;i}^{(1)} = L_{\bbH_i;i}^{(1)}$ and $L_{\bbH_j;i\neq j}^{(1)} = L_{\bbC;i}^{(1)}$. 

For the hybrid sampling $\bbH_j$, we have 
\begin{eqnarray}
L_{\bbH_j} &=& L_{\bbC}^{(0)} + 2\left(L_{\bbU;j}^{(1)} + \sum_{i\neq j} L_{\bbC;i}^{(1)}\right) \notag \\
&&+ O(\norm{F^{(1)}_i}^2,\norm{\Delta F}).
\end{eqnarray}
Then, the average of hybrid sampling is
\begin{eqnarray}
L_{\rm hybrid} &\equiv& \frac{1}{N_R} \sum_{i=1}^{N_R} L_{\bbH_i} \notag \\
&=& L_{\bbC}^{(0)} + 2\frac{1}{N_R}\sum_{i=1}^{N_R} \left[(N_R-1)L_{\bbC;i}^{(1)} + L_{\bbU;i}^{(1)}\right] \notag \\
&&+ O(\norm{F^{(1)}_i}^2,\norm{\Delta F}).
\end{eqnarray}
Note that the Clifford sampling gives 
\begin{eqnarray}
L_{\bbC} &=& L_{\bbC}^{(0)} + 2\sum_{i=1}^{N_R} L_{\bbC;i}^{(1)} + O(\norm{F^{(1)}_i}^2,\norm{\Delta F}).
\end{eqnarray}
We finally obtain the combined estimator of the error loss
\begin{eqnarray}
L_{\rm combined} &\equiv& N_R L_{\rm hybrid} - (N_R-1) L_{\bbC} \notag \\
&=& L_{\bbC}^{(0)} + 2\sum_{i=1}^{N_R} L_{\bbU;i}^{(1)} + O(\norm{F^{(1)}_i}^2,\norm{\Delta F}) \notag \\
&=& L_{\bbU}^{(0)} + 2\sum_{i=1}^{N_R} L_{\bbU;i}^{(1)} + O(\norm{F^{(1)}_i}^2,\norm{\Delta F}) \notag \\
&=& L_{\bbU} + O(\norm{F^{(1)}_i}^2,\norm{\Delta F}). 
\end{eqnarray}
This strategy can be readily generalized to higher orders, where stronger and higher-order correlated gate dependence can be tolerated by including more unitary single-qubit gates in the sampled circuits. In the discussion above we focus on the quadratic error loss function, but the same logic applies to the fidelity loss function as well.

\subsection*{Simulation of hybrid circuits}

To compute $L_{\bbH_j}$, we need to simulate error-free circuits with one non-Clifford single-qubit gate on a classical computer, which is efficient as discussed in Ref.~\cite{Bravyi2016}. A straightforward approach is to decompose a general unitary gate as a linear combination of ten linearly-independent Clifford gates~\cite{Endo2018}, i.e.~$[R] = \sum_{k=1}^{10} \alpha_k [B_k]$, where $B_k$ is a Clifford gate. Then, the error-free final state is a linear combination of ten stabiliser states, i.e.~$\rho^{\rm ef} = \sum_{k=1}^{10} \alpha_k \rho^{\rm ef}_k$. Here, $\rho^{\rm ef}_k$ is the final state of the Clifford circuit in which $R$ is replaced with $B_k$, and this circuit can be efficiently simulated using the classical computer. For the fidelity loss, we need to measure stabiliser operators of these ten stabiliser states in order to evaluate the fidelity in the non-stabiliser state $\rho^{\rm ef}$. 

\section{Numerical simulation}
\label{app:numerics}

\begin{figure}[tbp]
\centering
\includegraphics[width=1\linewidth]{\figpath 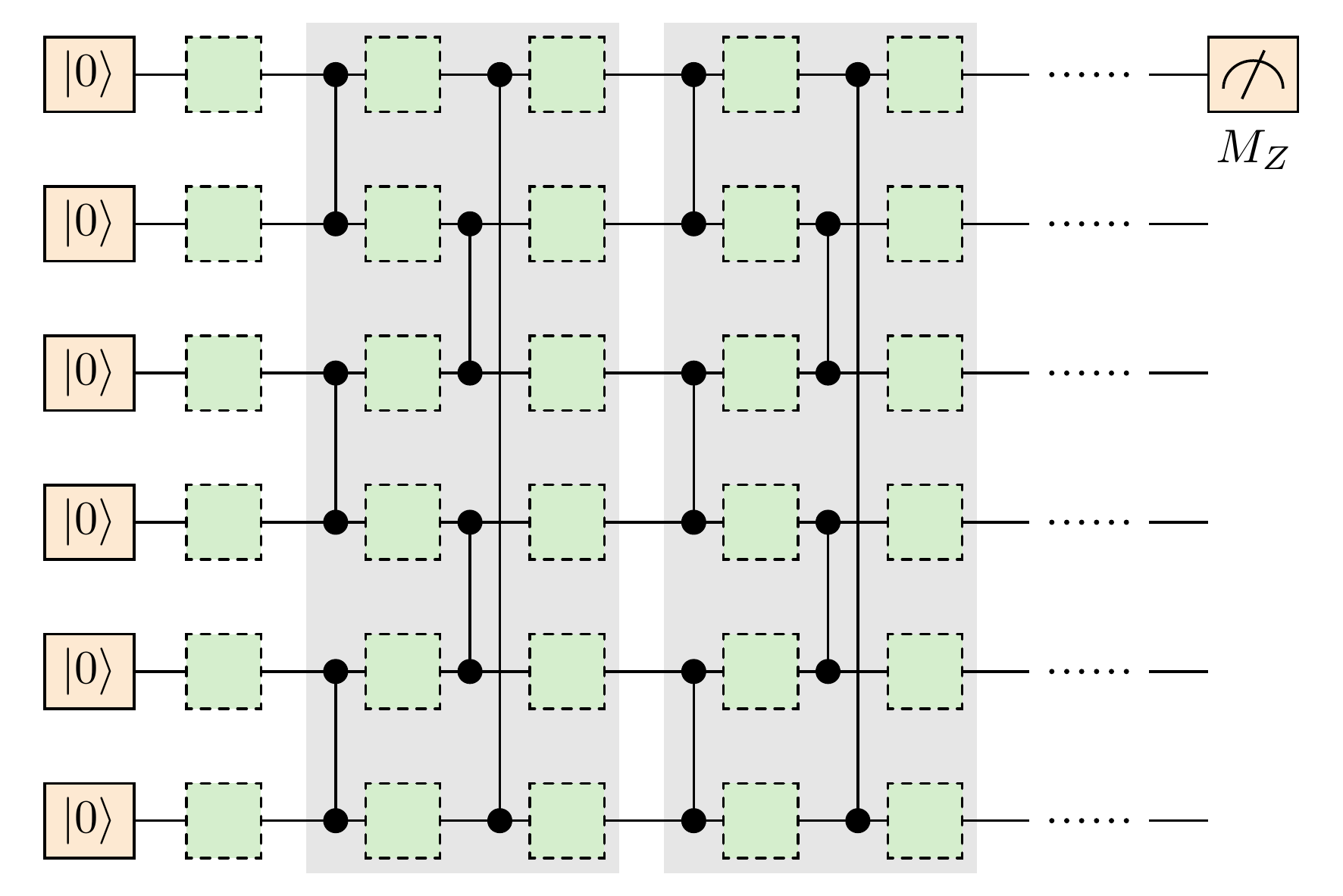}
\caption{
Circuit used in the numerical simulation. For a circuit of $n$ qubits, a layer of single-qubit gates contains one single-qubit gate on each qubit, and a layer of two-qubit gates has $n/2$ controlled-phase gates ($n$ is even). These $n$ qubits form a ring, i.e.~the $n$th qubit and the first qubit are coupled. Controlled-phase gates are applied on nearest-neighboring qubits. After the qubit initialization, a layer of single-qubit gates is applied. Then the circuit in the gray box is repeated for $n$ times, i.e.~the total number of controlled-phase gates is $n^2$. After the $n$th gray box, the first qubit is measured in the $Z$ basis. 
}
\label{fig:circuitStandard}
\end{figure}

Three categories of circuits are simulated using QuESTlink~\cite{Jones2020, Jones2019}. The first category includes circuits shown in Fig.~\ref{fig:circuitStandard}, and we call them standard circuits. The second category are randomly generated circuits, in which the observable is a tensor product of $Z$ operators on randomly selected qubits. The third category is the four-qubit circuit used in the experiment, see Fig.~3(d) in the main text. 

For the depolarizing, dephasing, amplitude damping, correlated coherent and gate-dependent depolarizing error models, we implement the simulation for the standard and randomly generated circuits with the qubit number four, six and ten. For each error model and qubit number, we take two different error rates. Given the error model, qubit number and error rate, we generate the standard circuit and three random circuits. Therefore, the total number of circuits is $5\times 3\times 2\times 4 = 120$. In the paper, we only show results of the standard circuit with $10$ qubits and one error rate for each error model. The complete data and codes for generating them are available at \href{https://github.com/yzchen-phy/clifford-sampling}{https://github.com/yzchen-phy/clifford-sampling}. 

For the composite error model and the experimentally-measured error model, we only implement the simulation for the four-qubit circuit used in the experiment. The hybrid sampling is demonstrated using the four-qubit standard circuit. 

In all the simulations, we assume qubit initialization and measurement are error-free. In the gate-dependent depolarizing error model (which is used in the hybrid sampling simulation), single-qubit gates are noisy. However, in all other error models, we assume single-qubit gates are also error-free. 

\subsection{Depolarizing model}

\begin{figure}[tbp]
\centering
\includegraphics[width=1\linewidth]{\figpath 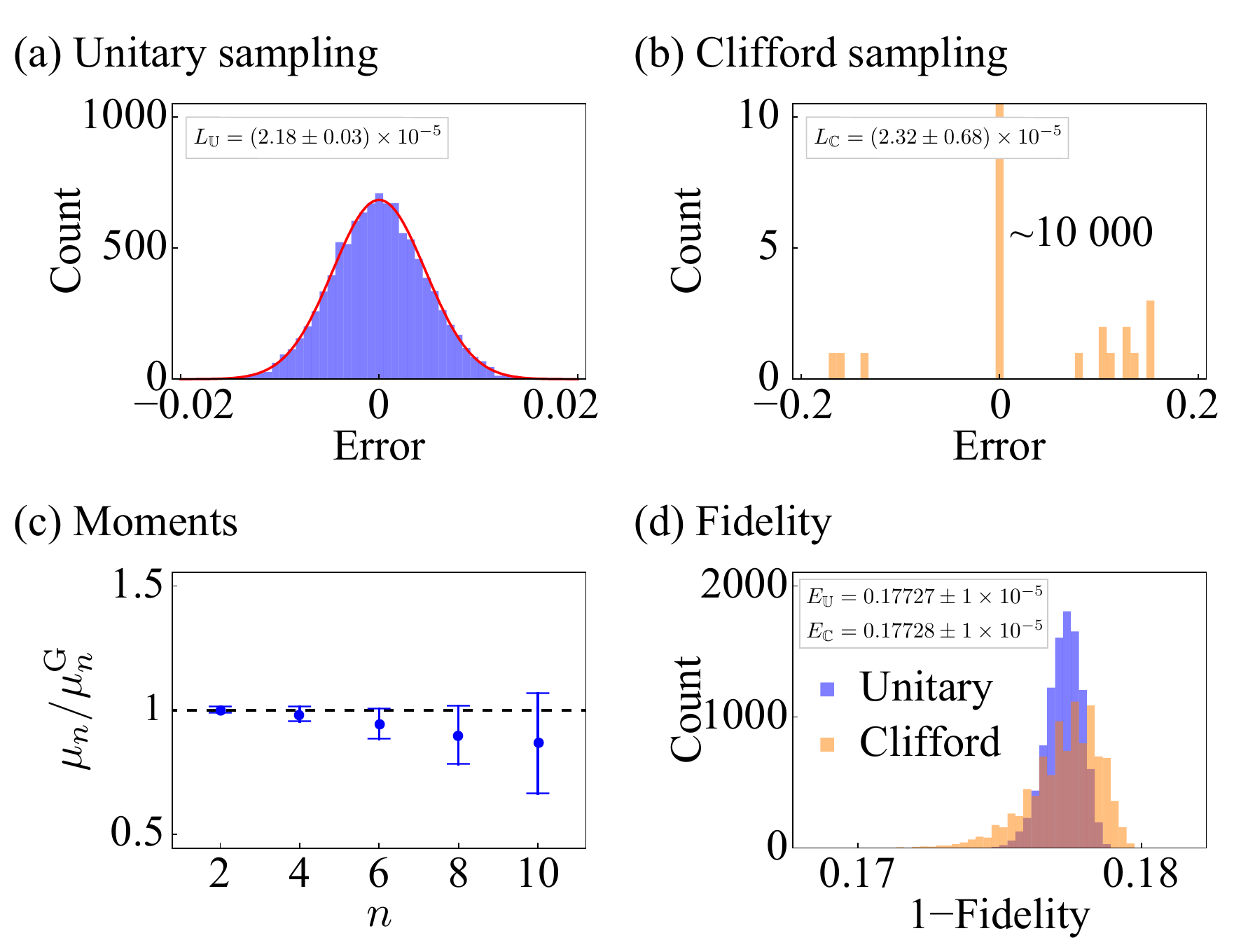}
\caption{
Numerical results of the depolarizing error model. 
}
\label{fig:Depol}
\end{figure}

We introduce the noise by adding the two-qubit depolarizing map after each two-qubit gate. The two-qubit depolarizing map reads
\begin{eqnarray}
\calN_2(\epsilon) = (1-\epsilon)[I] + \frac{\epsilon}{15}\sum_{\sigma\in\{I,X,Y,Z\}^{\otimes 2}\setminus I^{\otimes 2}}[\sigma].
\end{eqnarray}
The numerical results of the ten-qubit standard circuit with the error rate $\epsilon = 0.002$ are shown in Fig.~\ref{fig:Depol} (and Fig.~2 in the main text). Numbers of single-qubit-gate configurations are both $10 000$ in unitary sampling and Clifford sampling. 

\subsection{Dephasing model}

\begin{figure}[tbp]
\centering
\includegraphics[width=1\linewidth]{\figpath 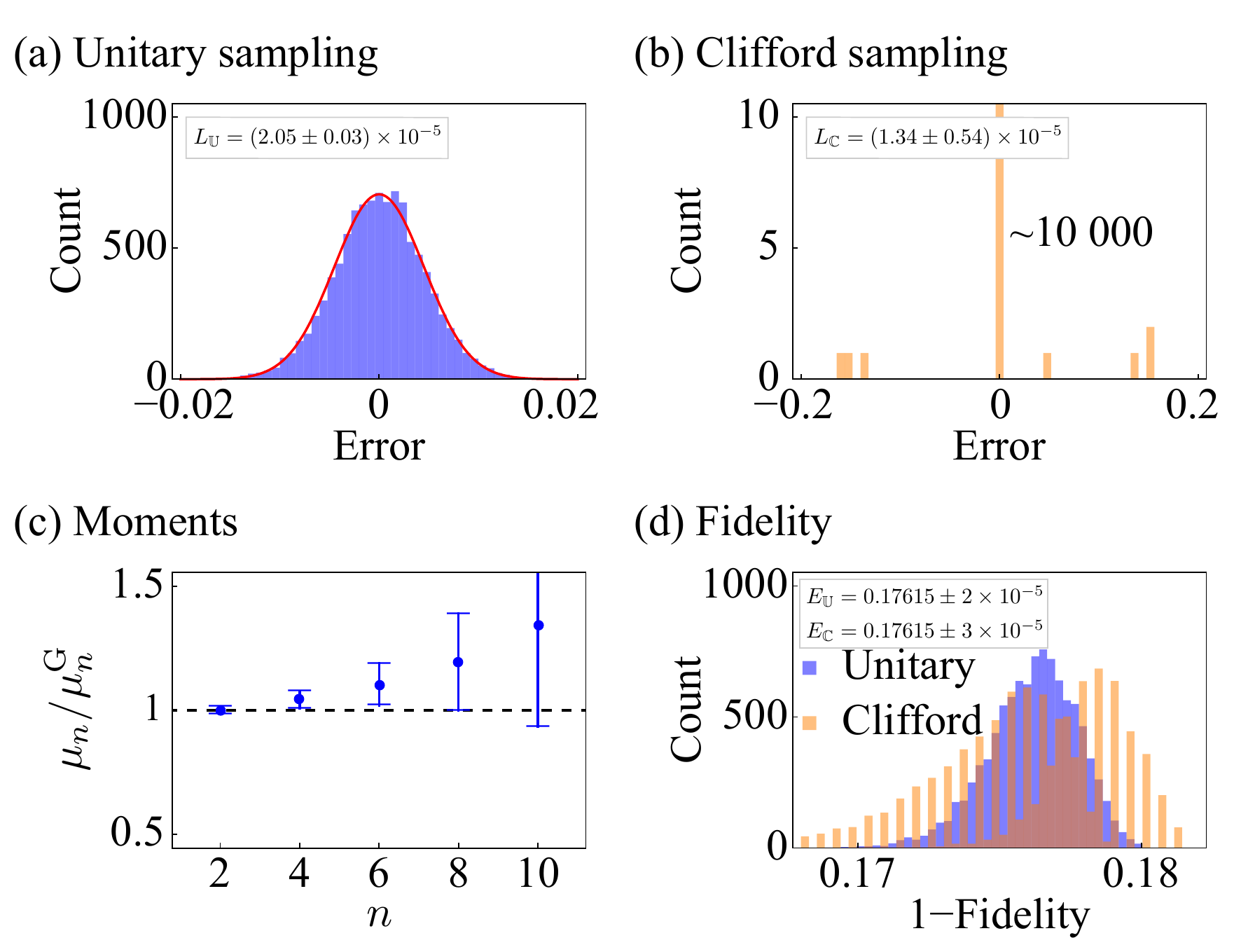}
\caption{
Numerical results of the dephasing error model. 
}
\label{fig:Deph}
\end{figure}

We introduce the noise by adding the two-qubit dephasing map after each two-qubit gate. The two-qubit dephasing map reads
\begin{eqnarray}
\mathcal{D}_2(\epsilon) = (1-\epsilon)[I] + \frac{\epsilon}{3}\sum_{\sigma\in\{I,Z\}^{\otimes 2}\setminus I^{\otimes 2}}[\sigma].
\end{eqnarray}
The numerical results of the ten-qubit standard circuit with the error rate $\epsilon = 0.002$ are shown in Fig.~\ref{fig:Deph}. Numbers of single-qubit-gate configurations are both $10 000$ in unitary sampling and Clifford sampling. 

\subsection{Amplitude damping model}

\begin{figure}[tbp]
\centering
\includegraphics[width=1\linewidth]{\figpath 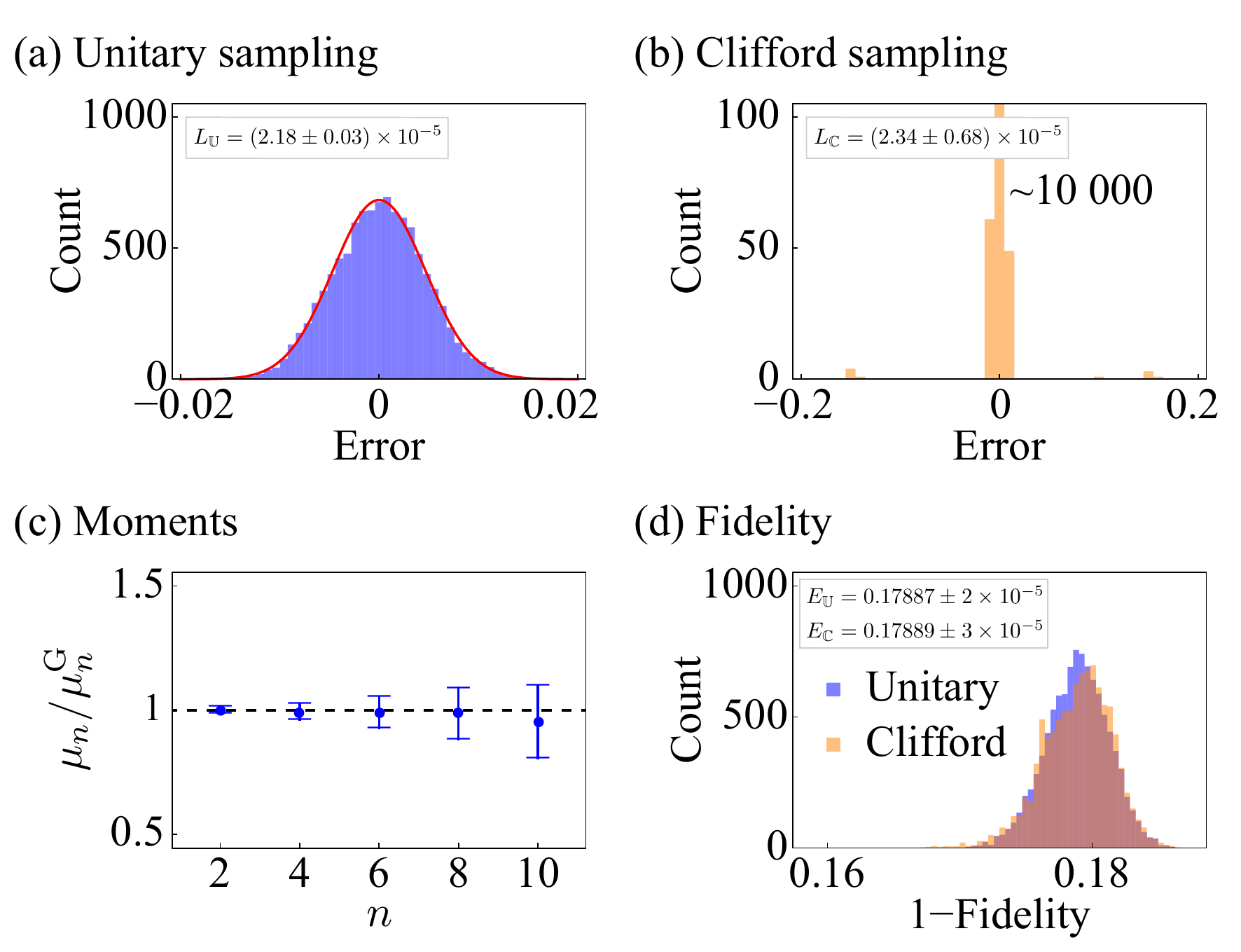}
\caption{
Numerical results of the amplitude damping error model. 
}
\label{fig:Damp}
\end{figure}

After each two-qubit gate, we introduce the noise by adding the one-qubit amplitude damping map on each qubit. The one-qubit amplitude damping map reads
\begin{eqnarray}
\mathcal{A}_1(\epsilon) &=& \left[\frac{I+Z}{2}+\sqrt{1-\epsilon}\frac{I-Z}{2}\right] \notag \\
&&+ \left[\sqrt{\epsilon}\frac{X+iY}{2}\right].
\end{eqnarray}
The numerical results of the ten-qubit standard circuit with the error rate $\epsilon = 0.002$ are shown in Fig.~\ref{fig:Damp}. Numbers of single-qubit-gate configurations are both $10 000$ in unitary sampling and Clifford sampling. 

\subsection{Correlated coherent model}

\begin{figure}[tbp]
\centering
\includegraphics[width=1\linewidth]{\figpath 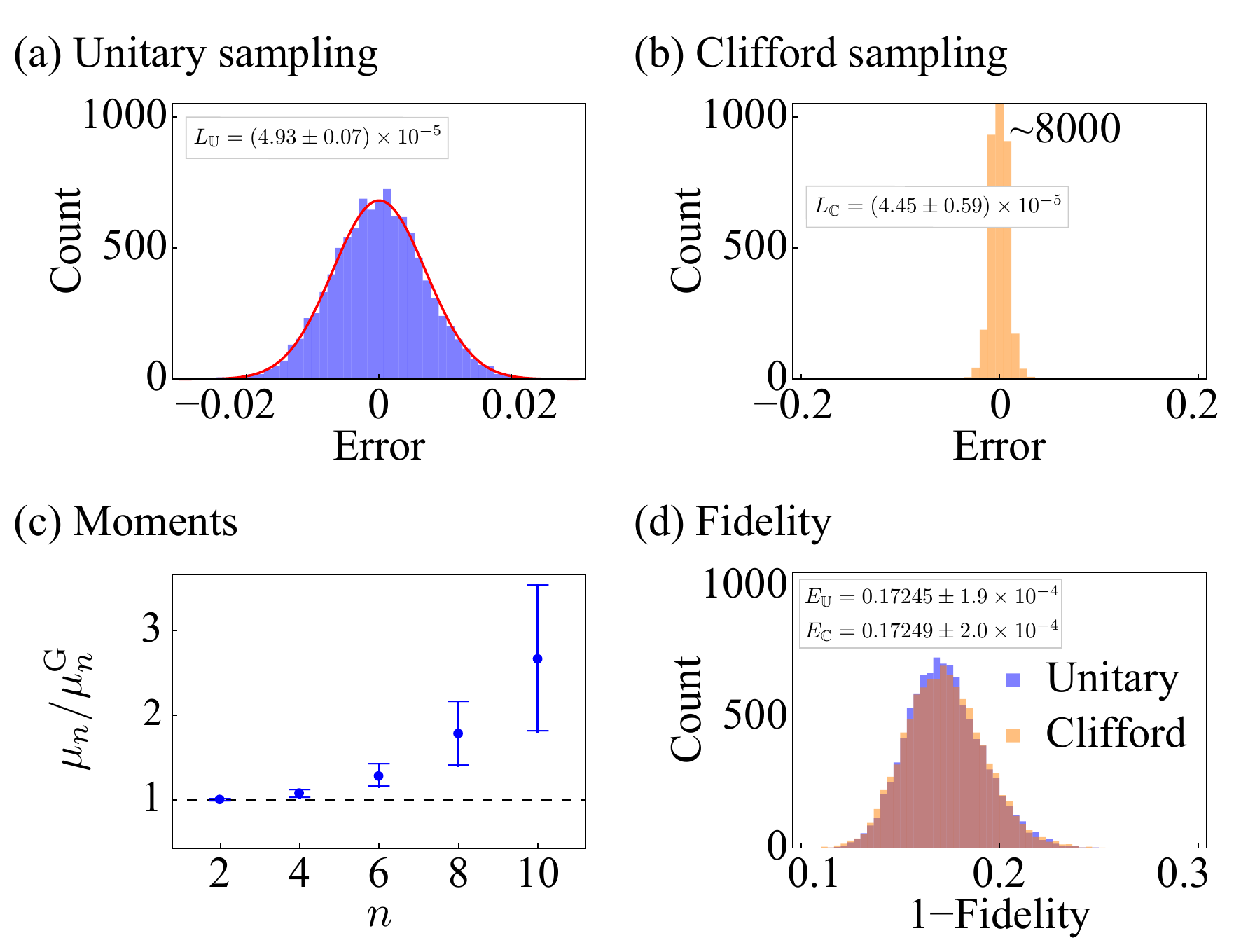}
\caption{
Numerical results of the correlated coherent error model. 
}
\label{fig:Coher}
\end{figure}

After each two-qubit gate, we introduce the noise by adding the one-qubit coherent-error map $[e^{\pm i \pi \epsilon Z}]$ on each qubit. In one run of the circuit, the sign is the same in all coherent-error maps, which is $+$ or $-$ with the probability of 1/2. The numerical results of the ten-qubit standard circuit with the error rate $\epsilon = 0.01$ are shown in Fig.~\ref{fig:Coher}. Numbers of single-qubit-gate configurations are both $10 000$ in unitary sampling and Clifford sampling. We can find that the high-order moments of the error distribution slowly deviate from those of the Gaussian distribution.

\subsection{Gate-dependent depolarizing model}

\begin{figure}[tbp]
\centering
\includegraphics[width=1\linewidth]{\figpath 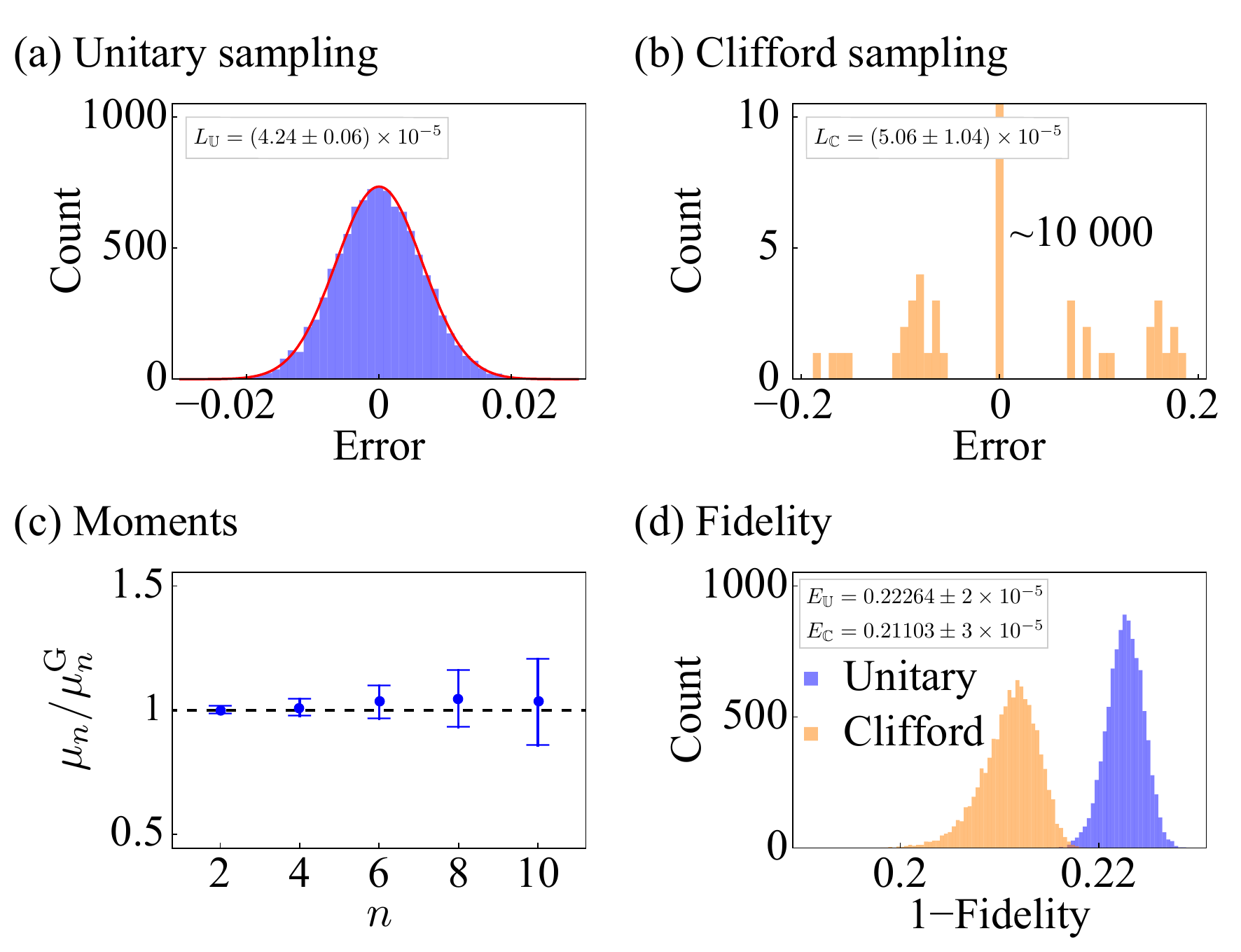}
\caption{
Numerical results of the gate-dependent depolarizing error model. 
}
\label{fig:SQ}
\end{figure}

The gate-dependent error model is based on the depolarizing error model. In addition to two-qubit depolarizing maps, we also introduce single-qubit depolarizing maps after each single-qubit gate. We decompose each single-qubit gate as $R = e^{-i \frac{\theta_3}{2} Z} e^{-i \frac{\theta_3}{2} X} e^{-i \frac{\theta_1}{2} Z}$. The single-qubit depolarizing map after the gate $R$ is $\calN_1(\gamma (\theta_1+\theta_2+\theta_3)/2\pi)$. The numerical results of the ten-qubit standard circuit with error rates $\epsilon = 0.002$ (two-qubit error rate) and $\gamma = \epsilon/10$ are shown in Fig.~\ref{fig:SQ}. Numbers of single-qubit-gate configurations are both $10 000$ in unitary sampling and Clifford sampling. 

\subsection{Composite model}

\begin{figure}[tbp]
\centering
\includegraphics[width=1\linewidth]{\figpath 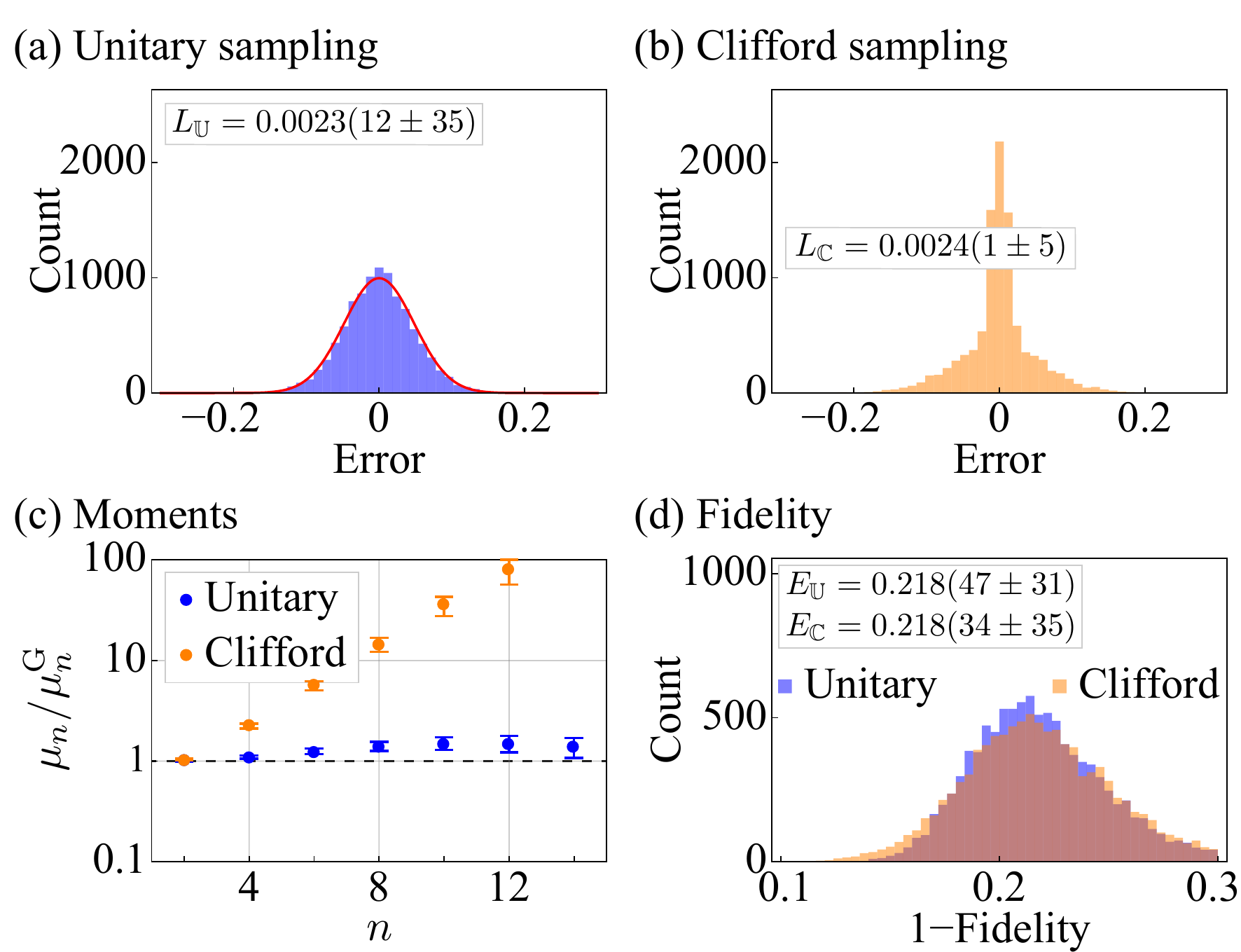}
\caption{
Numerical results of the composite error model. 
}
\label{fig:Tmodel}
\end{figure}

The composite error model consists of coherent single-qubit rotations and the amplitude damping map. After each two-qubit gate we introduce the single-qubit map $[e^{-i \pi \epsilon_X X}][e^{-i \pi \epsilon_Z Z}]$, followed by a single-qubit amplitude damping $\mathcal{A}_1(\epsilon_D)$. We set $\epsilon_D=0.02$ in the entire circuit while the values of $\epsilon_X$ and $\epsilon_Z$ are different for each two-qubit gate and drawn from the uniform distribution between $0$ and $0.04$. The noise parameters for the two qubits in a two-qubit gate are the same. The numerical results of the four-qubit experimental circuit are shown in Fig.~\ref{fig:Tmodel}. Numbers of single-qubit-gate configurations are both $10 000$ in unitary sampling and Clifford sampling. 

\subsection{Experimentally-measured model}

\begin{figure}[tbp]
\centering
\includegraphics[width=1\linewidth]{\figpath 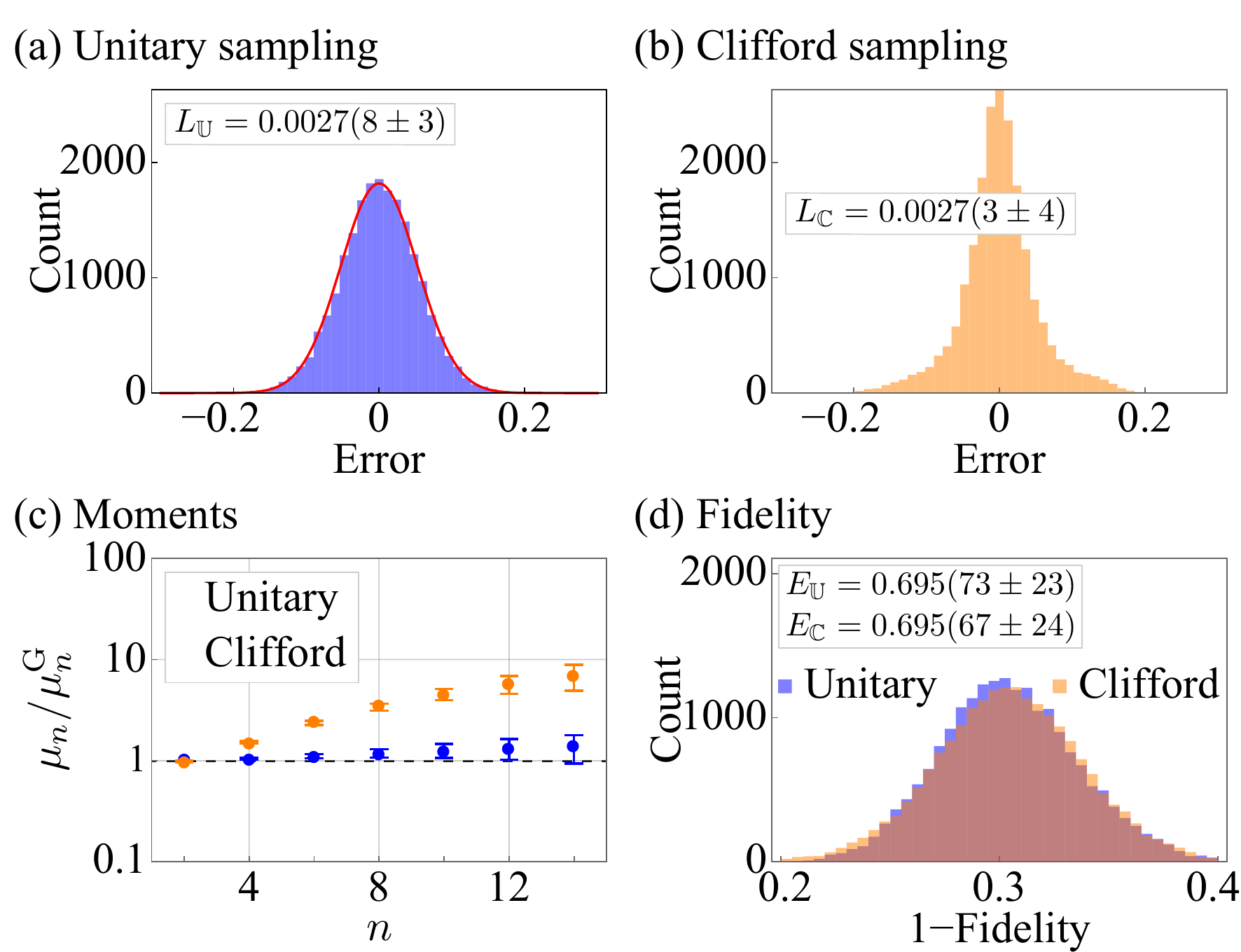}
\caption{
Numerical results of the experimentally-measured error model. 
}
\label{fig:Emodel}
\end{figure}

We simulate the four-qubit experimental circuit with the two-qubit gates replaced by the two-qubit maps obtained in the quantum process tomography (QPT). Other operations are assumed to be error-free. We note that the maps from QPT are not exactly trace-preserving and completely positive, as a result of sampling noise and state preparation and measurement error. Because QuESTlink validates maps, we prepared our own code to implement the numerical simulation without requiring the trace-preserving and completely positive condition. The numerical results are shown in Fig.~\ref{fig:Emodel}. Numbers of single-qubit-gate configurations are both $20 000$ in unitary sampling and Clifford sampling. 

\subsection{Hybrid sampling}

\begin{figure}[tbp]
\centering
\includegraphics[width=1\linewidth]{\figpath 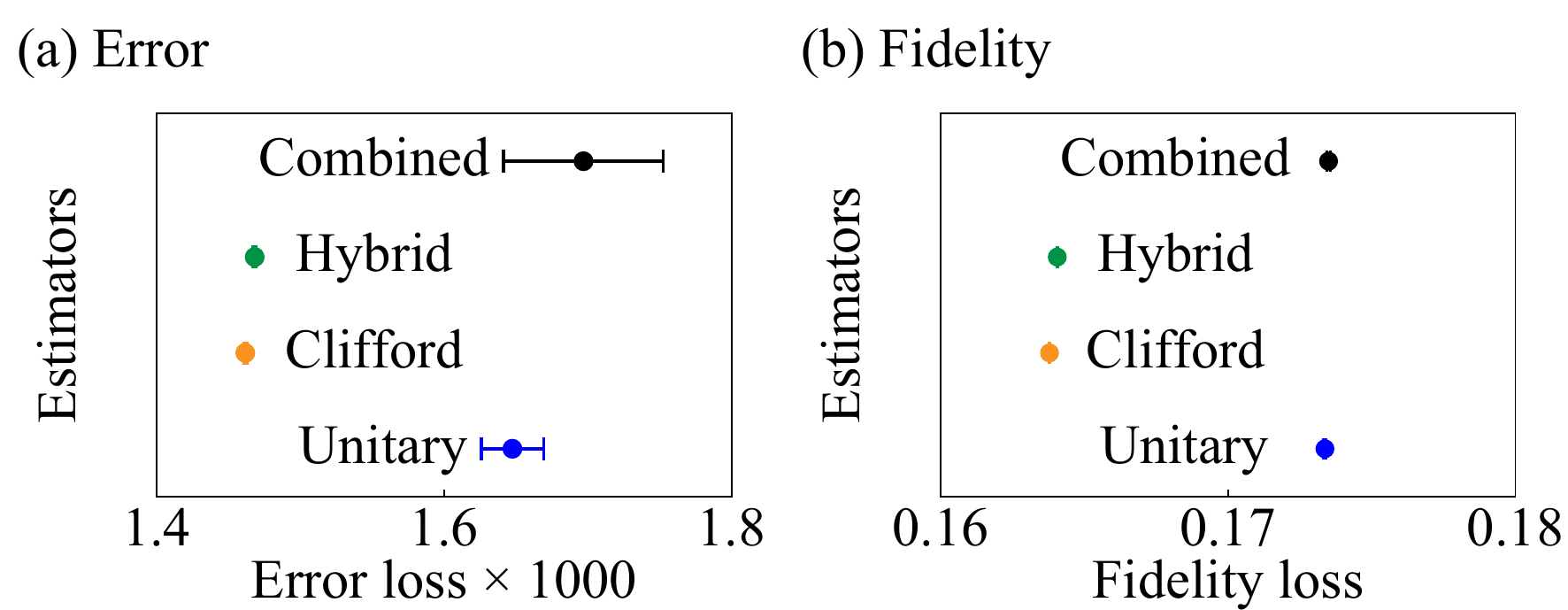}
\caption{
Numerical results of hybrid sampling. 
}
\label{fig:HS}
\end{figure}

We implement the simulation of the gate-dependent depolarizing model on the four-qubit standard circuit. We take $\epsilon = 0.01$ (two-qubit error rate) and $\gamma = \epsilon/10$. The number of single-qubit-gate configurations in the unitary sampling is $10 000$. The number of single-qubit-gate configurations in the Clifford sampling is $10 000\times 2(N_R-1)^2 = 4 500 000$. The number of single-qubit-gate configurations in the sampling of each $\bbH_i$ is $10 000\times 2N_R = 320 000$. Results are shown in Fig.~\ref{fig:HS}. We can find that the combined estimator is closer to the result of unitary sampling than fully-Clifford sampling. 

\section{Experimental details}
\label{app:exp}

\begin{table*}[tbp]
\caption{Quantum device parameters. $\omega_{j}$ is the idle frequency where we initialize the state of $Q_{j}$, apply single-qubit gates and perform quantum measurement. $g_{j}$ is the coupling strength between $Q_{j}$ and $B$. $T_{1, j}$ is the energy relaxation time and $T_{2, j}^*$ is the Ramsey (Gaussian) depahsing time of $Q_j$.  $F_{0, j}$ ($F_{1, j}$) is the readout fidelity for $Q_{j}$ in $\ket{0}$ ($\ket{1}$), which is used to eliminate the readout errors. $F_{R, j}^{\text{indiv}}$ is the RB gate fidelity of the $R$ gate, where $R\in \{X, Y, Z, X/2\}$, characterized individually for $Q_j$ at its idle frequency, whereas $F_{R, j}^{\text{simul}}$ is the simultaneous RB gate fidelity obtained by running RBs characterizing the same $R$ gate on all four qubits simultaneously \cite{2017_PRL_10GHZ}.}
\begin{center}
\begin{tabular}{p{3cm}<{\centering}p{3cm}<{\centering}p{3cm}<{\centering}p{3cm}<{\centering}p{3cm}<{\centering}}
  \hline
  \hline
                   Qubit    & $Q_{1}$   & $Q_{2}$   & $Q_{3}$   & $Q_{4}$ \\
  \hline
  $\omega_{j}/2\pi$ (\text{GHz})  & 4.904     & 4.955     & 4.999     & 5.043 \\

  $g_{j}/2\pi$ (\text{MHz})        & 18.9      & 17.5      & 16.1      & 18.9 \\

  $T_{1, j}$ (\text{$\mu$s})       & 38.7      & 54.4      & 34.3      & 37.4 \\

  $T_{2, j}^*$ (\text{$\mu$s})     & 2.2       & 2.3       & 2.2       & 2.5\\
  \hline
  $F_{0, j}$                & 0.957     & 0.981    & 0.977      & 0.973\\
  $F_{1, j}$                & 0.920     & 0.922    & 0.919      & 0.951\\
  \hline
  $F_{X, j}^{\text{indiv}}$           & 0.9983(2)    & 0.9967(4)    & 0.9993(1)    & 0.9991(1)\\
  $F_{X, j}^{\text{simul}}$           & 0.9986(6)    & 0.9966(17)    & 0.9936(15)    & 0.9995(7)\\
  $F_{Y, j}^{\text{indiv}}$           & 0.9987(2)    & 0.9978(3)    & 0.9993(1)    & 0.9994(1)\\
  $F_{Y, j}^{\text{simul}}$           & 0.9984(5)    & 0.9964(13)    & 0.9937(9)    & 0.9993(5)\\
  $F_{Z, j}^{\text{indiv}}$           & 0.9981(2)    & 0.9939(6)    & 0.9979(2)    & 0.9982(2)\\
  $F_{Z, j}^{\text{simul}}$           & 0.9982(8)    & 0.9975(12)    & 0.9954(11)    & 0.9984(5)\\
  $F_{X/2, j}^{\text{indiv}}$         & 0.9989(1)    & 0.9990(2)    & 0.9988(1)    & 0.9990(1)\\
  $F_{X/2, j}^{\text{simul}}$         & 0.9992(8)    & 0.9993(11)    & 0.9987(7)    & 0.9996(7)\\
  \hline
  \hline
\end{tabular}
\end{center}
\label{tab:deviceParameters}
\end{table*}

\subsection{Device parameters}

The quantum device consists of 20 frequency-tunable Xmon qubits, where four qubits labeled as $Q_{1} \sim Q_{4}$ are used in this experiment to illustrate the idea of Clifford sampling. Each qubit has a Z line for tuning the qubit frequency, an XY line for exciting the qubit and a readout resonator coupled to a common readout line for the qubit-state measurement. With regard to connectivity, each qubit is capacitively coupled to the central bus resonator $B$, which has a fixed resonance frequency of $\omega_{\text{B}}/2\pi \approx 5.248$ \text{GHz}, with the coupling strength ($g_{j}$) listed in Tab.~\ref{tab:deviceParameters}. In the experiment, $Q_{1} \sim Q_{4}$ are initialized in the ground state $\ket{0}$ at their respective idle frequencies $\omega_{j}$ as listed in Tab.~\ref{tab:deviceParameters}, while all the other qubits are left at their respective maximum frequencies, i.e., the sweetpoints insensitive to flux noises, which are at least 1 \text{GHz} higher than the idle frequencies of $Q_{1} \sim Q_{4}$. The energy relaxation time ($T_{1, j}$) and the Ramsey dephasing time ($T_{2, j}^*$) are listed in Tab.~\ref{tab:deviceParameters}. It was observed that the dressed states of the qubits under coherent microwave fields are less sensitive to external dephasing noises~\cite{Guo2018}. Therefore the effective dephasing times of these qubits should be much longer than $T_{2, j}^*$, as we frequently apply the $XY$ microwave driving fields on these qubits to implement the single-qubit rotations, the single-qubit dynamical decoupling schemes and the two-qubit $U_{\text{phase}}$ gates within the Clifford sampling circuit. 
 
\subsection{Readout correction}

\begin{table*}[tbp]
\caption{Two-qubit gate parameters. $\omega_{\text{I}, jj'}$ is the interaction frequency at which we tune $Q_{j}$ and $Q_{j'}$ on resonance to obtain $U_{\text{phase}, jj'}$. 
$\lambda_{jj'}$ is the effective coupling strength between $Q_{j}$ and $Q_{j'}$.
By applying continuous driving fields with driving strengths of $\Omega_j$ and $\Omega_j'$ on $Q_{j}$ and $Q_{j'}$, respectively, for a fixed duration of $T_{\text{gate}, jj'}$, we realize $U_{\text{phase}, jj'}$. The individual (simultaneous) gate fidelity $F$ of $U_{\text{phase}, jj'}$ is obtained by QPT on $Q_{j}$-$Q_{j'}$, while no action ($U_{\text{phase}, kk'}$ in parallel) is applied on the other two qubits $Q_{k}$-$Q_{k'}$ at the idle frequencies ($\omega_{\text{I}, kk'}$).}
\begin{center}
\begin{tabular}{p{5cm}<{\centering}p{2cm}<{\centering}p{2cm}<{\centering}p{2cm}<{\centering}p{2cm}<{\centering}p{2cm}<{\centering}p{2cm}<{\centering}}
  \hline
  \hline
   Qubit pair       & $Q_{1}Q_{2}$ & $Q_{1}Q_{3}$ & $Q_{1}Q_{4}$ & $Q_{2}Q_{3}$ & $Q_{2}Q_{4}$ & $Q_{3}Q_{4}$ \\
  \hline
   $\omega_{\text{I}, jj'}/2\pi$ (\text{GHz}) & 4.905   & 4.905   & 4.955  & 4.955  & 5.043 & 5.043 \\
  
   $\lambda_{jj'}/2\pi$ (\text{MHz})   & 0.90    & 0.95    & 0.85   & 0.70   & 1.1   & 0.80 \\
	$\Omega_{j(j')}/2\pi$ (\text{MHz}) & 10.93(2.96)   &3.61(12.74)     & 2.96(9.09)   & 2.96(13.39)   & 10.13(4.26)  & 7.52(12.74)\\
 
  $T_{\text{gate}, jj'}$ (\text{ns})          & 277     & 274     & 297    & 356    & 231   & 326 \\

  $F$($U_{\text{phase}, jj'}$) (Individual)& 0.967(2)   &  0.968(3)   & 0.968(3)    & 0.944(3)  & 0.951(4) & 0.960(2) \\

  $F$($U_{\text{phase}, jj'}$) (Simultaneous) & 0.956(2)   &  0.961(4)   & 0.941(3)    & 0.941(3)  & 0.947(5) & 0.962(2) \\
  \hline
  \hline
\end{tabular}
\end{center}
\label{tab:two qubit gate}
\end{table*}

The experimental scheme to directly measure the multiqubit occupation probabilities and the subsequent procedure to eliminate the readout errors were detailed in Ref.~\cite{2017_PRL_10GHZ,2019_Science_20qubit}. 
To initialize the qubit in its ground state $\ket{0}$, we idle it for about 200~$\mu$s, during which the residue thermal excitation is estimated to be small via a postselection procedure. The produced $\ket{0}$ state of the qubit has a state fidelity above 0.99 on average, following which we can apply a high-fidelity $X$ gate ($\pi$ rotation around $x$-axis of the Bloch sphere) to reliably prepare the qubit in $\ket{1}$.

However, due to the existence of readout errors, the directly measured occupation probability $P_{0}^{m}$ ($P_{1}^{m}$) after we reliably prepare $Q_{j}$ in $\ket{0}$ ($\ket{1}$) may still be away from ideal (actual), which is noted as $Q_{j}$'s measurement fidelity in $\ket{0}$ ($\ket{1}$), $F_{0, j}$ ($F_{1, j}$). 
For any experimental measurement, we label the directly measured probability vector as $\bfP^{m}=(P_{0}^m, P_{1}^m)^T$ and the actual one as $\bfP=(P_{0}, P_{1})^T$, 
which map as
\begin{eqnarray}
\bfP_{j} =\begin{pmatrix}F_{0, j} & 1 - F_{1, j}\\1-F_{0, j} & F_{1, j}\end{pmatrix}^{-1}\bfP_{j}^m=\bfF_{j}^{-1}\bfP_{j}^{m}.
\end{eqnarray}
To eliminate the readout errors of the directly measured probability column vector $\bfP_{Q_1Q_2Q_3Q_4}^{m}$ for the 4-qubit joint states, we obtain the actual probability column vector by
\begin{eqnarray}
\bfP_{Q_1Q_2Q_3Q_4}=(\otimes_{j=1}^{4}\bfF_{j})^{-1}\bfP_{Q_1Q_2Q_3Q_4}^{m}.
\end{eqnarray}
Table~\ref{tab:deviceParameters} lists the simultaneously obtained measurement fidelity values $F_{0,j}$ and $F_{1,j}$ for all four qubits.

\subsection{Single-qubit gate}

\begin{figure*}[tbp]
\centering
\includegraphics[width=1\linewidth]{\figpath 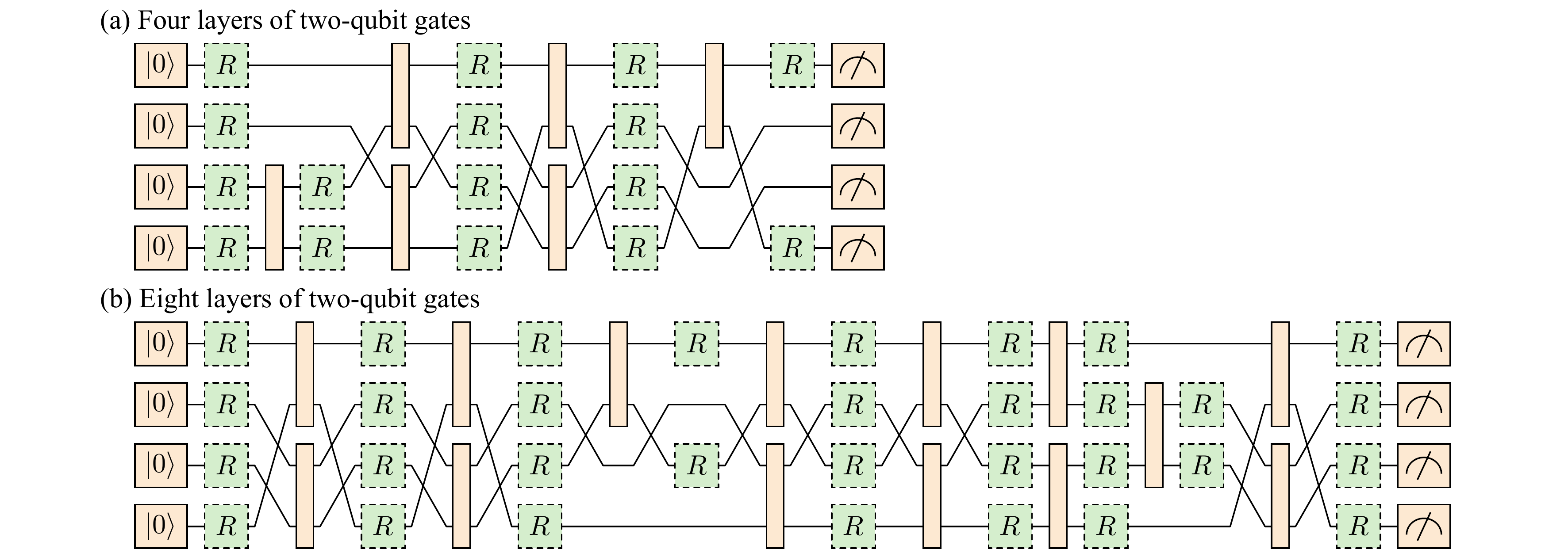}
\caption{Exemplary frame operation configurations. (a) Configuration with four layers of two-qubit gates. (b) Configuration with eight layers of two-qubit gates.}
\label{fig:twoFrames}
\end{figure*}

Single-qubit gates include unitary single-qubit gates and Clifford single-qubit gates. 
For a unitary single-qubit gate, we randomly generate a unitary matrix $M_U\in {\rm U}(2)$, which is distributed with Haar measure~\cite{2020SciPy-NMeth}. For a Clifford single-qubit gate, we randomly choose a matrix $M_C\in {\rm C}(2)$. Here, ${\rm C}(2)$ denotes the Clifford group of one qubit. Then we convert the matrix to $e^{i\alpha}R_{xy}(\theta_{xy}, \phi_{xy})R_{z}(\theta_{z})$, where $R_{z}(\theta_z)$ represents a rotation by $\theta_{z}$ around $z$-axis and $R_{xy}(\theta_{xy}, \phi_{xy})$ represents a rotation by $\theta_{xy}$ around the axis in the equator plane, which has an angle $\phi_{xy}$ with respect to $x$-axis.  
Here the global phase factor $e^{i\alpha}$ can be ignored. To characterize the gate performance, we perform individual and simultaneous randomized benchmarkings (RBs) on respresentative single-qubit gates such as the $X$ ($\theta_{xy}= \pi, \phi_{xy}=0$), $X/2$ ($\theta_{xy}= \pi/2, \phi_{xy}=0$), $Y$ ($\theta_{xy}= \pi, \phi_{xy}= \pi/2$) and $Z$ ($\theta_z = \pi$) gates, yielding gate fidelities no less than 0.993 for all four qubits (see Tab.~\ref{tab:deviceParameters}).

\subsection{Two-qubit $U_{\text{phase}}$ gate}

\begin{figure}[tbp]
\centering
\includegraphics[width=1\linewidth]{\figpath/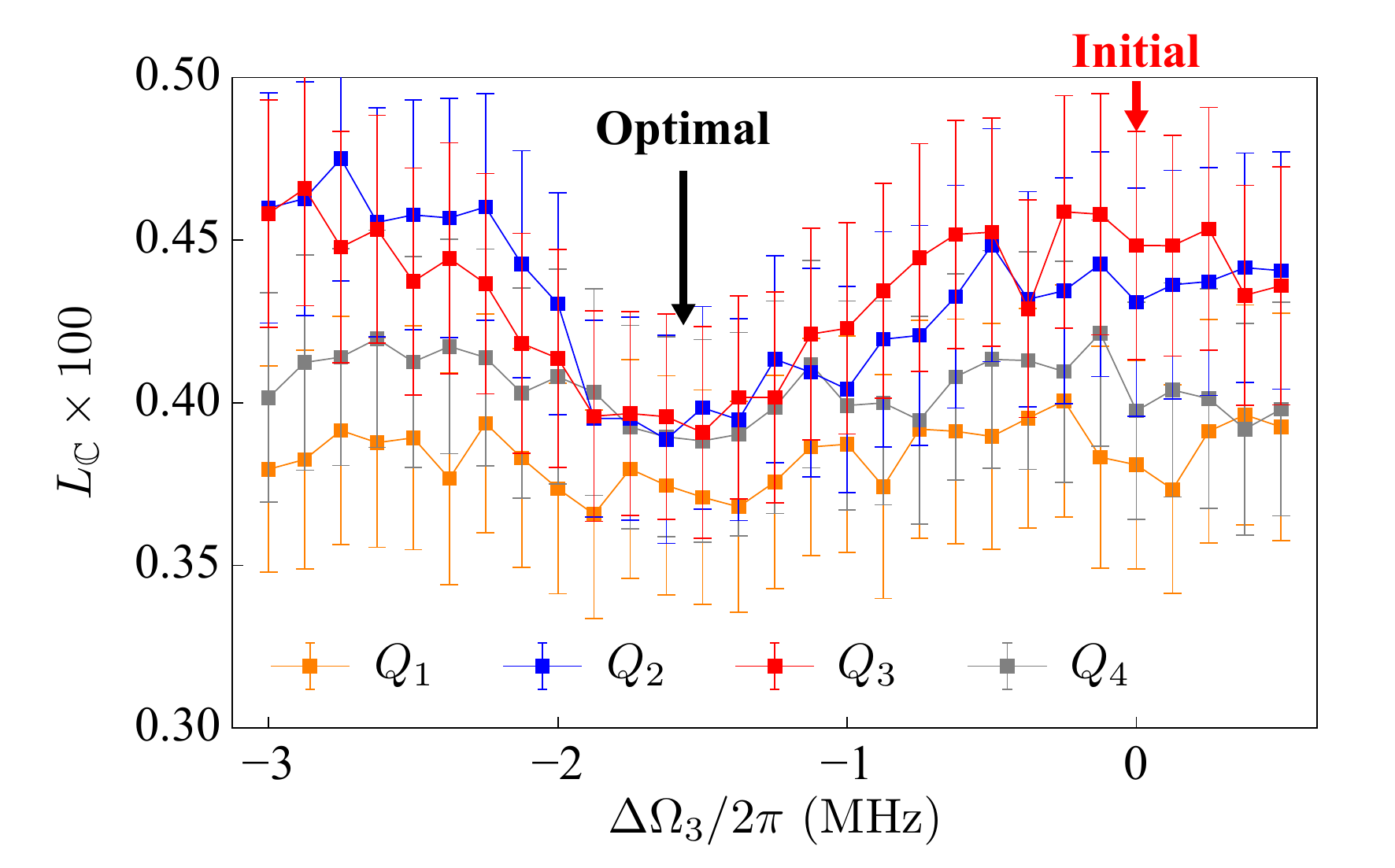}
\caption{Parameter optimization of the $U_{\text{phase}}$ gate on $Q_2$-$Q_3$ based on Clifford sampling.}
\label{fig:Optimisation}
\end{figure}

The $U_{\text{phase}}$ gate implemented on a pair of qubits was detailed in Ref.~\cite{Guo2018}. Here we discuss the case of executing parallel $U_{\text{phase}}$ gates on four qubits. The original Hamiltonian with four qubits is
\begin{eqnarray}
H /\hbar &=& \omega_B a^{\dagger}a + \sum_{j=1}^{4} \omega_{j}|1_j\rangle \langle 1_j|\notag\\
&&+\sum_{j=1}^{4} \left[ g_{j}(\sigma _{j}^{+}a+\sigma _{j}^{-}a^{\dagger})\right],  \label{eq1}
\end{eqnarray}
where $\sigma _{j}^{+}$ ($\sigma _{j}^{-}$) is the raising (lowering) operator of $Q_{j}$ and $a^{\dagger}$ ($a$) is the creation (annihilation) operator of $B$.

Now we divide the four qubits into two pairs, $Q_k$-$Q_{k'}$ and $Q_{l}$-$Q_{l'}$, and position $Q_k$-$Q_{k'}$ ($Q_{l}$-$Q_{l'}$ ) at the interaction frequency $\omega_{\text{I}, kk'}$ ($\omega_{\text{I}, ll'}$). Under the assumption that $\omega_{\text{I}, kk'}$ and $\omega_{\text{I}, ll'}$ are largely detuned and $B$ is in its ground state, we have
\begin{eqnarray}
H_1 /\hbar &=& \sum_{j\in\{k,l\}}\lambda_{jj'}(\sigma_{j}^{-}\sigma_{j'}^{+}+\sigma_{j}^{+}\sigma_{j'}^{-})\notag\\
&&+\sum_{j\in\{k,l\}}(\frac{g_{j}^2}{\Delta_{jj'}}|1_{j}\rangle \langle 1_{j}|+\frac{g_{j'}^2}{\Delta_{jj'}}|1_{j'}\rangle \langle 1_{j'}|),
\end{eqnarray}
where $\Delta_{jj'}=\omega_{I, jj'}-\omega_{B}$ and $\lambda_{jj'}$ is effective coupling strength between $Q_{j}$ and $Q_{j'}$.
Therefore the intra-pair qubits interact with each other, while the inter-pair qubits are effectively decoupled. 
By applying appropriate microwave driving fields on all four qubits, the Hamiltonian in the double-rotating frames can be written as:
\begin{eqnarray}
H_{\text{eff}}/\hbar &=& \sum_{j\in\{k,l\}}(\lambda_{jj'}\sigma_{j}^{+}\sigma_{j'}^{-}+\Omega_{j}e^{-i\phi_{j}}\sigma_{j}^++\Omega_{j'}e^{-i\phi_{j'}}\sigma_{j'}^+)\notag\\
&&+\text{H.c.},
\label{hamiltonian}
\end{eqnarray}
where $\Omega_{j}$ ($\phi_{j}$) is the Rabi frequency (phase) of the driving field on $Q_{j}$. The Hamiltonian for the two qubits ($Q_{j}$-$Q_{j'}$) in a pair takes exactly the form as described in Eq.~(4) of Ref.~\cite{Guo2018}, based on which a controlled $\pi$-phase gate in the dressed-state basis can be realized by choosing appropriate driving strengths and evolving the system for a fixed amount of time $T_{\text{gate}, jj'}$, during which the phases of the driving fields are reversed at $T_{\text{gate}, jj'}$/2.
The unitary matrix in the two-qubit computational basis $\{\ket{00}, \ket{01}, \ket{10}, \ket{11}\}$ describing the evolution is 
\begin{equation}
U_{\text{phase}}=\frac{1}{\sqrt{2}}\begin{pmatrix}
1 & 0 & 0 & i \\
0 & 1 & i & 0 \\
0 & i & 1 & 0 \\
i & 0 & 0 & 1
\end{pmatrix}.
\end{equation} 
The resulting $U_{\text{phase}}$ gates on various qubit combinations are characterized by QPT and summarized in Tab.~\ref{tab:two qubit gate}.

\subsection{Randomly generated frame operation configurations}

\begin{figure}[tbp]
\centering
\includegraphics[width=1\linewidth]{\figpath 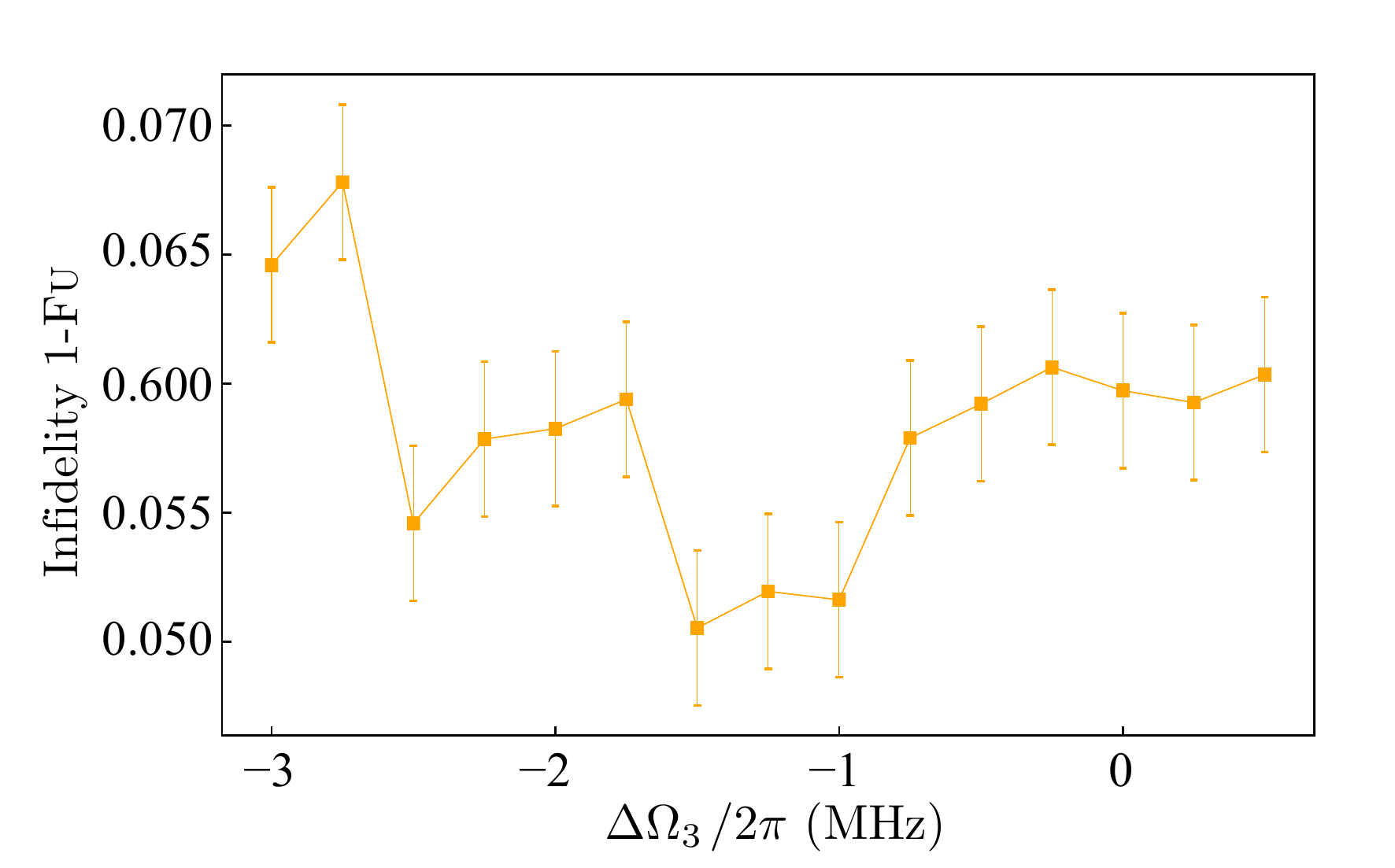}
\caption{
Infidelity of the $U_{\text{phase}}$ gate on $Q_2$-$Q_3$ characterized by QPT as a function of $\Delta\Omega_{3}$.
}
\label{fig:optFid}
\end{figure}

As mentioned in the main text, we randomly generate 50 frame operation configurations to validate $L_{\bbU}$= $L_{\bbC}$. Here we plot two examples in Fig.~\ref{fig:twoFrames}: one with four layers and a total of six $U_{\text{phase}}$ gates [Fig.~\ref{fig:twoFrames}(a)], and the other with eight layers and a total of fourteen $U_{\text{phase}}$ gates [Fig.~\ref{fig:twoFrames}(b)]. For the four-layer example, there are two $U_{\text{phase}}$ gates in each of the middle layers and one $U_{\text{phase}}$ gate in each of the beginning and ending layers. 
For any layer with only one $U_{\text{phase}}$ gate, during the $U_{\text{phase}}$ gate on two qubits for a time of about 300 \text{ns} (see Tab.~\ref{tab:two qubit gate}), we apply continuous microwave fields, with the driving strengths $\Omega/2\pi\approx 3$ \text{MHz}, on the other two qubits which are left at their idle frequencies to protect them from dephasing~\cite{Guo2018}. 

\subsection{Clifford sampling optimization}

Usually, pulse parameters of a gate are predetermined by benchmarking qubits individually. Nonetheless, the gate performance may decline, or the optimal parameters may drift when we implement multiple gates in parallel, because of correlations such as crosstalk. The more qubits are involved, the more significant the impact of correlations may become. Clifford sampling provides a convenient and scalable way to optimize parameters in the large-circuit quantum computation. As a demonstration, we detect the optimal parameter of the gate $U_{\text{phase}}$ on qubits $Q_2$ and $Q_3$ in the experiment. An important parameter of $U_{\text{phase}}$ is the strength of driving field, e.g.~Rabi frequency $\Omega_{2}$ ($\Omega_{3}$) applied on $Q_2$ ($Q_3$). In the experiment, we find the optimal $\Omega_{3}$ in the four-qubit circuit shown in Fig.~3(d) in the main text. The observable is $E_f = \ketbra{0}{0}_i$, i.e.~the probability in $\ket{0}$ for one of four qubits. We change $\Omega_{3}$ from its initial value {$\Omega_{3}^{\rm ini} = 2\pi\times13.39\text{ MHz}$} to $\Omega_{3}^{\rm ini} + \Delta\Omega_{3}$. For each value of $\Omega_{3}$, we implement $1500$ random single-qubit gate configurations $\bfR$ to estimate the Clifford error loss $L_{\bbC}$. The result is shown in Fig.~\ref{fig:Optimisation}. We can find that error losses of $Q_2$ and $Q_3$ decrease by more than $10\%$ when $\Delta\Omega_{3}$ changes from $0$ to $-2\pi\times 1.5\text{ MHz}$. 

To demonstrate the effectiveness of the $L_{\bbC}$ approach, we measure the infidelity of the $Q_2$-$Q_3$ $U_{\text{phase}}$ gate, $1-F_{U}$, as a function of $\Delta{\Omega_{3}}$, while we apply the $U_{\text{phase}}$ gates on $Q_2$-$Q_3$ and $Q_1$-$Q_4$ simultaneously. As shown in Fig.~\ref{fig:optFid}, the infidelity approaches the minimum around $\Delta\Omega_{3} = -2\pi\times1.5$ MHz, agreeing well with the Clifford sampling optimization, proving the effectiveness of optimizing with $L_{\bbC}$ as the loss function. 

\noindent{$^*$ Z. W. and Y. C. contributed equally to this work.\\
$^\dagger$ chaosong@zju.edu.cn\\
$^\ddagger$ yli@gscaep.ac.cn}

\bibliography{CliffordSamp}

\begin{thebibliography}{63}
\bibitem [{foo()}]{footnote}%
  \BibitemOpen
  \href@noop {} {}\bibinfo {note} {We observe that the mean is approximately
  zero in all numerical simulations and experiments, in which the observable is
  a Pauli operator. In general, the mean can be measured using Clifford
  sampling and Hybrid sampling in the same way as the loss
  function.}\BibitemShut {Stop}%
\end{thebibliography}

\end{document}